\documentclass{aa}  

\usepackage{graphicx}
\usepackage{tabularx}
\usepackage{multicol}
\usepackage{txfonts}
\usepackage{booktabs}
\usepackage{mathtools}
\usepackage[colorlinks=true, allcolors=blue]{hyperref} 
\usepackage[position=top]{subcaption}

\usepackage{tikz}
\usetikzlibrary{shapes.geometric, arrows, positioning}

\usepackage[capitalise]{cleveref}
\Crefname{figure}{Figure}{Figures} 
\crefname{figure}{Fig.}{Figs.} 
\Crefname{tabular}{Table}{Tables}
\crefname{tabular}{Table}{Tables} 
\Crefname{section}{Section}{Sections}
\crefname{section}{Sect.}{Sects.}

\newcommand\mathbold[1]{\boldsymbol{#1}}
\newcommand\mathboldit[1]{\mathbold{#1}}

\newcommand\LCDM{\text{$\Lambda$CDM}}

\newcommand\COLA{\mathrm{COLA}}
\newcommand\GR{\mathrm{GR}}
\newcommand\BD{\mathrm{BD}}
\newcommand\MG{\mathrm{MG}}
\newcommand\lin{\mathrm{lin}}
\newcommand\prim{\mathrm{prim}}

\newcommand\params{\mathbold{\theta}}
\newcommand\hiclass{\textsc{hi\_class}}
\newcommand\hmcode{\textsc{hmcode}}
\newcommand\halofit{\textsc{halofit}}
\newcommand\fml{\textsc{fml}}
\newcommand\ramses{\textsc{ramses}}
\newcommand\eetwo{\textsc{EuclidEmulator2}}
\newcommand\bdpy{\textsc{bd.py}}

\begin{document}


% ...
\title{A simple prediction of the nonlinear matter power spectrum in Brans-Dicke gravity from linear theory}

\author{Herman Sletmoen \and Hans A. Winther}

\institute{Institute of Theoretical Astrophysics, University of Oslo, P.O.Box 1029 Blindern, N-0315 Oslo, Norway\\
          \email{herman.sletmoen@astro.uio.no}
         }

 
\abstract{
Brans-Dicke (BD), one of the first proposed scalar-tensor theories of gravity, effectively makes the gravitational constant of general relativity (GR) time-dependent.
Constraints on the BD parameter $\omega$ serve as a benchmark for testing GR, which is recovered in the limit $\omega \rightarrow \infty$.
Current small-scale astrophysical constraints $\omega \gtrsim 10^5$ are much tighter than large-scale cosmological constraints $\omega \gtrsim 10^3$,
but the two decouple if the true theory of gravity features screening.
On the largest cosmological scales, BD approximates the most general second-order scalar-tensor (Horndeski) theory, so constraints here have wider implications.
These constraints will improve with upcoming large-scale structure and cosmic microwave background surveys.
To constrain BD with weak gravitational lensing, one needs its nonlinear matter power spectrum $P_\BD$.
By comparing the boost $B = P_\BD/P_\GR$
from linear theory and nonlinear $N$-body simulations,
we show that the nonlinear boost can simply be predicted from linear theory
if the BD and GR universes are parameterized in a way that makes their early cosmological evolution and quasilinear power today similar.
In particular, they need the same $H_0 / \sqrt{\smash[b]{G_{\rm eff}(a=0)}}$ and $\sigma_8$,
where $G_{\rm eff}$ is the (effective) gravitational strength.
Our prediction is $1\%$ accurate for $\omega \geq 100$, $z \leq 3$, and $k \leq 1\,h/\mathrm{Mpc}$; and $2\%$  up to $k \leq 5\,h/\mathrm{Mpc}$.
It also holds for $G_\BD$ that do not match Newton's constant today,
so one can study GR with different gravitational constants $G_\GR$ by sending $\omega \rightarrow \infty$.
We provide a code that computes $B$ with the linear Einstein-Boltzmann solver \hiclass{} and multiplies it by the nonlinear $P_\GR$ from \eetwo{} to predict $P_\BD$. 
}

\authorrunning{H. Sletmoen and H. A. Winther}
\titlerunning{A simple prediction of the nonlinear matter power spectrum in Brans-Dicke gravity}
\maketitle

\section{Introduction}
\label{sec:intro}

The Brans-Dicke (BD) theory \citep{bransMachPrincipleRelativistic1961a,dickeMachPrincipleInvariance1962a} of modified gravity \citep[MG;][]{clifton_modified_2012} features a scalar field in addition to the spacetime metric.
It effectively promotes Newton's constant in general relativity (GR) to a dynamical gravitational strength.
This scalar-tensor theory was introduced in the 1960s to implement Mach's principle in GR.
Since then it has become perhaps the most famous alternative to Einstein's theory and a root for the development of other theories,
such as the more general Horndeski theory \citep{horndeski_second-order_1974}.
BD features an additional parameter $\omega$ and reduces to GR as $\omega \rightarrow \infty$,
so BD can be understood as a generalization of GR.
In cosmology, replacing GR with BD in the standard $\Lambda$ cold dark matter ((GR-)$\LCDM$) model gives the alternative BD-$\LCDM$ model,
where the effective gravitational strength evolves with time and modifies the expansion history and growth of structure.

\begin{table*}[]
\footnotesize
\centering
\caption{Constraints on the BD parameter $\omega$ from various tests, beginning with the Cassini bound from 2003.}
\begin{tabularx}{\textwidth}{X r r}
\toprule
Experiment (dataset) & Constraint $\downarrow$ (conf.) & Year \\ 
\midrule
Pulsar timing (PSR J0337+1715) \citep{voisinImprovedTestStrong2020} & $\omega > 140\,000$ ($95\%$) & 2020 \\
Pulsar timing (PSR J0337+1715) \citep{archibald_testing_2018} & $\omega > \phantom{0}73\,000$ ($95\%$) & 2018 \\
Shapiro time delay (Cassini satellite) \citep{bertottiTestGeneralRelativity2003,willConfrontationGeneralRelativity2014} & $\omega > \phantom{0}40\,000$ ($95\%$) & 2003 \\ 
Pulsar timing (PSR J1738+0333) \citep{freireRelativisticPulsarwhiteDwarf2012} & $\omega > \phantom{0}25\,000$ ($68\%$) & 2012 \\
\midrule
CMB (\textit{Planck}), BAO (\textit{BOSS}, \textit{6dF}, \textit{SDSS}) \citep{Ooba_2017} & $\omega > \phantom{00}2\,000$ ($95\%$) & 2017 \\
CMB (\textit{Planck}), BAO (\textit{BOSS}, \textit{6dF}, \textit{SDSS}), Hubble (several) \citep{amirhashchiConstrainingExactBransDicke2020} & $\omega > \phantom{00}1\,560$ ($95\%$) & 2019 \\
CMB (\textit{Planck}),  BAO (\textit{BOSS}), SN (\textit{Pantheon}), LSS (\textit{KiDS}, \textit{2dFLenS}) \citep{joudakiTestingGravityCosmic2022} & $\omega > \phantom{00}1\,540$ ($95\%$) & 2022 \\ 
CMB (\textit{Planck}) \citep{avilezCosmologicalConstraintsBransDicke2014} & $\omega > \phantom{000\,}890$ ($99\%$) & 2014 \\ 
CMB (\textit{Planck}), BAO (\textit{6dF}, \textit{SDSS}) \citep{Umilt__2015} & $\omega > \phantom{000\,}208$ ($95\%$) & 2015 \\
CMB (\textit{WMAP}, \textit{Planck}), BAO (\textit{SDSS}, \textit{6dF}) \citep{liConstraintsBransDickeGravity2013} & $\omega > \phantom{000\,}182$ ($95\%$) & 2013 \\
CMB (\textit{WMAP}, others), LSS (\textit{2dF}) \citep{Acquaviva_2005} & $\omega > \phantom{000\,}120$ ($95\%$) & 2005 \\
CMB (\textit{WMAP}, others), LSS (\textit{SDSS}) \citep{wu_cosmic_2010,Wu_2010} & $\omega > \phantom{000\,0}98$ ($95\%$) & 2010 \\
\bottomrule
\end{tabularx}
\tablefoot{Sorted by $\omega$, the table naturally divides into two ``camps'' from small-scale and large-scale constraints. We exclude $\omega < 0$ due to ghost instability. Note that \cite{avilezCosmologicalConstraintsBransDicke2014} and \cite{joudakiTestingGravityCosmic2022} report separate constraints with and without fixing $G_{m0} = G$ today; we only show the constraints when it is not fixed, which results in the tightest constraints. Also note that \cite{joudakiTestingGravityCosmic2022} report different bounds $\omega > \{1540,160,160,350\}$ depending on the prior used for $\omega$ in the Markov chains.
See also \cite{Nagata_2004,Ooba_2016,Ooba_2017,Ballardini_2016,Ballardini_2020,Umilt__2015} for constraints on the gravitational strength and generalized BD models.}
\label{tab:constraints}
\end{table*}

As summarized in \cref{tab:constraints}, observations in the $21$st century have placed very strong constraints on $\omega$.
Already in 2003, Shapiro time delay measurements of the Cassini satellite on its way to Saturn  constrained $\omega \gtrsim 10^4$,
and recent strong-field tests from timing rapidly rotating neutron stars (pulsars) even bound $\omega \gtrsim 10^5$.
On cosmological scales, however, the story is different.
Roughly speaking, constraints have tightened from only $\omega \gtrsim 10^2$ with stage-II survey data in the 2000s to $\omega \gtrsim 10^3$ with stage-III surveys in the 2010s.
In the next decade, \cite{alonso_observational_2017} expects cosmological constraints to improve by another order of magnitude with upcoming stage-IV surveys like \textit{Euclid} \citep{laureijsEuclidDefinitionStudy2011}, \textit{DESI} \citep{desi_collaboration_desi_2016}, the \textit{Vera C. Rubin Observatory} \citep{lsst_science_collaboration_lsst_2009}, \textit{SKA} \citep{yahya_cosmological_2015} and next-generation cosmic microwave background (CMB) experiments \citep{abazajian_cmb-s4_2016}.
Fisher forecasts found that if $\omega = 800$,
then \textit{Euclid} should be able to constrain this to $3\%$ using galaxy clustering and weak lensing data \citep{fruscianteEuclidConstrainingLinearly}.
Likewise, \cite{Ballardini_2019} forecast that upcoming clustering and weak lensing data in combination with \textit{BOSS} and CMB observations have the potential to reach $\omega \gtrsim 10^4$ in the most optimistic case.
With these seemingly ever-tightening competing constraints, BD has become a benchmark theory for testing (deviations from) GR.
BD also survived the burial of theories with gravitational wave speeds $v \neq c$ from GW170817 \citep{ezquiaga_dark_2017},
and \cite{solaBransDickeGravityCosmological2019,solaBransDickeCosmologyLambda2020} show that it alleviates the $H_0$ and $S_8$ tensions \citep{perivolaropoulosChallengesCDMUpdate2022a}. 

In fact, several scalar-tensor theories reduce to BD on large (cosmological) scales, where gradients of the scalar field are suppressed   \citep{avilezCosmologicalConstraintsBransDicke2014,joudakiTestingGravityCosmic2022}.
This makes BD interesting not only on its own, but also as the large-scale limit of more general theories, so constraints on it have wider implications. 
Such theories typically feature screening mechanisms that hide the scalar field in dense small-scale regions, letting GR take over there.
In other words, the true theory of gravity could be similar to BD on the largest scales, similar to GR on the smallest scales, and transition between them over intermediate scales.
The very tight small-scale (astrophysical) constraints could therefore be irrelevant from a cosmological perspective.

In this work, we aim to develop a tool for predicting the nonlinear BD-$\LCDM$ matter power spectrum $P_\BD(k \lesssim 5\,h/\mathrm{Mpc},\, z \lesssim 3 \,|\, \params)$ across parameter space $\params$ to around $1\%$ precision, for use in constraining the theory with upcoming stage-IV large-scale structure survey data.
There are several possible approaches to this, all of which rely on $N$-body simulations to account for the nonlinear evolution.

\textbf{Halo modeling tools} like \hmcode{} \citep{mead_accurate_2015,mead_hmcode-2020:_2021} and \halofit{} \citep{smith_stable_2003,takahashi_revising_2012} rely on dark matter's clustering in halos to effectively reverse-engineer the nonlinear matter distribution from halo statistics. They can be integrated directly with linear Einstein-Boltzmann solvers. The \hmcode{} framework has been extended to account for a wide range of dark energy and MG models, massive neutrinos and baryonic physics \citep{mead_accurate_2016}.
Another related method is the halo model reaction approach of \cite{Cataneo_2019}, which combines the halo model and perturbation theory to model corrections coming from nonstandard physics.
This is implemented in the publicly available \textsc{ReACT} code \citep{2020MNRAS.498.4650B,Giblin_2019}.
The halo modeling approach was taken in \cite{joudakiTestingGravityCosmic2022},
who used $N$-body simulations to create a fitting formula in \hmcode{} to predict the nonlinear BD-$\LCDM$ power spectrum.
As far as we know, this is the only nonlinear prediction for BD so far.

\textbf{Emulators} are trained from a set of (time-consuming) $N$-body simulations for a limited number of cosmological parameters, and then rely on machine learning to (quickly) interpolate in parameter space and output the nonlinear matter power spectrum for arbitrary cosmologies.
There are already sophisticated emulators for the GR-$\LCDM$ power spectrum, such as \eetwo{} \citep{euclidcollaborationEuclidPreparationIX2021}, \textsc{CosmicEmu} \citep{heitmannMiraTitanUniversePrecision2016,lawrenceMiraTitanUniverseII2017,moranMiraTitanUniverseIV2023}, and \textsc{BACCO} \citep{2021MNRAS.507.5869A,2021MNRAS.506.4070A,2021arXiv210414568A}.
Recent works, including \cite{wintherEmulatorsNonlinearMatter2019}, \cite{brandoEnablingMatterPower2022}, \cite{fiorini_fast_2023}, \textsc{Sesame} \citep{maulandSesamePowerSpectrum2023}, and \textsc{e-MANTIS} \citep{saez-casares_e-mantis_2023}, have produced emulators for selected MG theories. It is also possible to combine halo modeling with emulation and emulate the ingredients of the halo model \citep{ruan2023emulatorbased}.
At the time of writing, no dedicated BD-$\LCDM$ power spectrum emulator exists.

\textbf{Simulation rescaling} \citep{angulo_one_2010} is a technique where data from a single $N$-body simulation is rescaled to produce output for a different cosmology.
This technique was in fact used in training the \textsc{BACCO} emulator.

We demonstrate a hybrid technique, somewhat reminiscent of both rescaling and emulation, suited to extending existing predictors of GR-$\LCDM$ to BD-$\LCDM$.
By carefully selecting the cosmological parameters $\params_\BD$ and $\params_\GR$,
we map a given BD universe to a corresponding GR universe such that their cosmological evolutions are as similar as possible.
This makes it possible to predict the nonlinear matter power spectrum boost $B = P_\BD/P_\GR \approx P_\BD^\lin/P_\GR^\lin$ using linear power spectra from cheap Einstein-Boltzmann solvers, which we verified by comparing it to the boost obtained from $N$-body simulations.
In turn, we can predict the full BD-$\LCDM$ power spectrum $P_\BD = B \cdot P_\GR$ by combining it with any existing high-quality emulator for $P_\GR$.
This ``trick'' exploits the effort that has already gone into creating sophisticated GR emulators,
and avoids duplicating this work for every alternative model to be explored.
It could be a viable cheap route for constraining MG theories with upcoming surveys
\citep{fruscianteEuclidConstrainingLinearly,casas_linear_2017}. 

The rest of this paper is structured as follows.
\Cref{sec:bd} reviews the background cosmology and perturbation theory of BD (and thus GR).
\Cref{sec:pipeline} describes our pipeline for the Einstein-Boltzmann solver and $N$-body simulations.
\Cref{sec:parametrization} justifies the parameterization we use to compute the boost.
\Cref{sec:results} shows the resulting linear and nonlinear boost and compares them to older predictions.
\Cref{sec:conclusions} concludes.

\section{Brans-Dicke modified gravity}\label{sec:bd}

Brans-Dicke (BD) theory \citep{bransMachPrincipleRelativistic1961a,dickeMachPrincipleInvariance1962a} generalizes general relativity (GR).
In addition to the metric tensor $g_{\mu\nu}$,
it features a scalar field $\phi$ with a constant dimensionless parameter $\omega$.
Here we use units where $c = 1$,
but explicitly keep Newton's constant $G=6.67 \cdot 10^{-11}\,\mathrm{Nm}^2/\mathrm{kg}^2$ to show how BD effectively modifies the gravitational constant of GR.
In the Jordan frame,
the total action of BD gravity minimally coupled to the matter action $S_M$ is
\begin{equation}
S\left[g_{\mu\nu},\phi\right] = \frac{1}{16 \pi G} \int \mathrm{d}^4 x \sqrt{-|g|} \, \bigg[ \phi R - \frac{\omega}{\phi} (\nabla \phi)^2 \bigg] + S_M,
\label{eq:action}
\end{equation}
where $g_{\mu\nu}$ is the spacetime metric
with determinant $|g|<0$, Ricci scalar $R$ and covariant derivative $\nabla_\mu$;
and $(\nabla \phi)^2 = (\nabla^\mu \phi) (\nabla_\mu \phi) = g^{\mu\nu} (\nabla_\mu \phi) (\nabla_\nu \phi)$.
From the principle of least action $\delta S = 0$,
the (classical) equations of motion for $g_{\mu\nu}$ and $\phi$ are
the modified Einstein and Klein-Gordon field equations
\citep{clifton_modified_2012}
\begin{subequations}
\begin{align}
G_{\mu\nu} &= \frac{8 \pi G}{\phi} T_{\mu\nu} + \frac{\omega}{\phi^2} \bigg[ (\nabla_\mu \phi) (\nabla_\nu \phi) - \frac12 g_{\mu\nu} (\nabla \phi)^2 \bigg] \nonumber\\&+ \frac{1}{\phi} \bigg[ \nabla_\mu \nabla_\nu \phi - g_{\mu\nu} \nabla^2 \phi \bigg], \label{eq:eom1} \\
\nabla^2 \phi &= \frac{8 \pi G}{3+2\omega} T, \label{eq:eom2}
\end{align}%
\label{eq:eom}%
\end{subequations}%
where $\nabla^2 = g^{\mu\nu} \nabla_\mu \nabla_\nu$ is the four-dimensional Laplacian,
and $T_{\mu\nu} = (-2/\sqrt{-g}) \, (\delta S_M / \delta g^{\mu\nu})$ is the energy-momentum tensor of matter with trace $T = T^\mu_{\phantom{\mu}\mu}$.
Effectively, $\phi$ promotes Newton's constant $G$ in the Einstein field equations \eqref{eq:eom1} to a dynamical field $G/\phi$.
In the following, we see in more detail how BD is similar to GR with such an effective gravitational strength.

GR is recovered in the limit $\omega \rightarrow \infty$:
then $\nabla^2 \phi = 0$,
whose solution $\phi = 1$ everywhere recovers Einstein's field equations
$G_{\mu\nu} = 8 \pi G \, T_{\mu\nu}$ with Newton's constant $G$.

\subsection{Cosmology}

We assume a spatially flat, homogeneous and isotropic Friedmann-Lemaître-Robertson-Walker background with small perturbations in the Newtonian gauge:
\begin{equation}
\mathrm{d}s^2 = -\big[1+2\Psi(t,\mathbf{x})\big] \mathrm{d}t^2 + a^2(t) \big[1+2\Phi(t,\mathbf{x})\big] \mathrm{d}\mathbf{x}^2.
\label{eq:flrw}
\end{equation}
We fill the universe with a perfect fluid with total energy-momentum tensor
\begin{equation}
    T^{\mu\nu} = \sum_s T^{\mu\nu}_{(s)} = \sum_s (\rho_s+P_s) u_s^\mu u_s^\nu + P_s g^{\mu\nu},
\end{equation}
with energy densities $\rho_s$, pressures $P_s$, and four-velocities $u^\mu_s$ from the species $s$:
\begin{itemize}
\item radiation $r$ (from photons $\gamma$ and massless\footnote{Including massive neutrinos is straightforward, but previous studies (such as \cite{maulandSesamePowerSpectrum2023}) have shown that they have little effect on the matter power spectrum boost.} neutrinos $\nu$) with equation of state $\bar{P}_r = \bar\rho_r/3$,
\item matter $m$ (from baryons $b$ and cold dark matter $c$) with equation of state $\bar{P}_m = 0$,
\item dark energy $\Lambda$ (from a cosmological constant) with equation of state $\bar{P}_\Lambda = -\bar\rho_\Lambda$.%
\end{itemize}
The universe's background and (linear) perturbations are described perturbatively up to first order around a homogeneous and isotropic universe.

\subsection{Background cosmology}
\label{sec:background}

To zeroth order, the equation of motions \eqref{eq:eom}  give the Friedmann and scalar field equations
\begin{subequations}
\begin{align}
H^2 &= \frac{8 \pi G}{3 \bar\phi} \bar\rho + \frac{\omega}{6} \left(\frac{\dot{\bar\phi}}{\bar\phi}\right)^2-H\frac{\dot{\bar\phi}}{\bar\phi}, \quad  ({}_{\mu\nu} = {}_{00}) \label{eq:friedmann_dimensionful} \\
\frac{1}{a^3} \frac{\mathrm{d}}{\mathrm{d}t} \left( \dot{\bar\phi} a^3 \right) &= \frac{8 \pi G}{3+2\omega} (\bar\rho - 3\bar{P}), \label{eq:scalar_field_dimensionful}
\end{align}%
\label{eq:background_dimensionful}%
\end{subequations}%
where $\bar\rho = \bar\rho_r + \bar\rho_m + \bar\rho_\Lambda$,
$\bar{P} = \bar{P}_r + \bar{P}_m + \bar{P}_\Lambda$,
$\dot{{}} = \mathrm{d}/\mathrm{d}t$ and $H = \dot{a}/a$ is the Hubble parameter.
The Friedmann equation in BD is the same as in GR,
only with additional effective energy from $\smash{\dot{\bar\phi}}$
and an effective gravitational strength $G_H = G/\bar\phi$.

According to the Boltzmann equation, the three species evolve as
$\bar\rho_r = \bar\rho_{r0} a^{-4}$, $\bar\rho_m = \bar\rho_{m0} a^{-3}$ and $\bar\rho_\Lambda = \bar\rho_{\Lambda 0} a^{0}$, where ${}_0$ indexes quantities by their values today.
Defining the standard density parameters $\Omega_s = \bar\rho_s / (3 H^2/8 \pi G)$,
physical densities today are measured by $\omega_{s0} = \Omega_{s0} h^2 \propto \bar\rho_{s0}$.%
\footnote{Some authors instead define $\Omega_s = \rho_s / (3 H^2 / 8 \pi (G/\bar\phi))$ in terms of the effective $G/\bar\phi$. This is natural in BD because their sum is $\sum_s \Omega_s = 1$ at all times, if one defines an effective $\Omega_\phi$. However, we opted for the GR definition $\Omega_s = \rho_s / (3 H^2 / 8 \pi G)$ to minimize confusion and maintain consistency with the \hiclass{} input parameter $\texttt{Omega\_cdm} = \rho_s / (3 H_0^2 / 8 \pi G)$.}
Re-parameterizing the evolution in terms of ${}^\prime = \mathrm{d}/\mathrm{d} \ln a = H^{-1} \cdot \mathrm{d}/\mathrm{d}t$,
the system \eqref{eq:background_dimensionful} can be rewritten in the dimensionless form
\begin{subequations}
\begin{align}
E^2 = \left(\frac{H}{H_0}\right)^2 &= \frac{1}{\bar\phi} \cdot \frac{\displaystyle \Omega_{r0} a^{-4} + \Omega_{m0} a^{-3} + \Omega_{\Lambda 0}}{1 - \frac{\omega}{6} [(\ln\bar\phi)^\prime]^2 + (\ln\bar\phi)^\prime}, \label{eq:friedmann_dimensionless} \\
\big(\bar\phi^\prime E \, a^3 \big)^\prime &= \frac{3}{3+2\omega}\frac{\Omega_{m0} + 4 \, \Omega_{\Lambda 0} \, a^{-3}}{E} .
\label{eq:background_dimensionless}
\end{align}
\end{subequations}

This system can be integrated from an initial scale factor $a_\mathrm{ini}$,
given $\bar\phi(a_\mathrm{ini})$, $\bar\phi^\prime(a_\mathrm{ini})$, $\Omega_{r0}$, $\Omega_{m0}$, $\Omega_{\Lambda 0}$, and $\omega$.
But these are not all free parameters,
as today's Friedmann equation $E_0 = 1$ constrains one of them -- say $\Omega_{\Lambda 0}$.
To find a consistent solution, one can iterate over $\Omega_{\Lambda 0}$
and solve the system repeatedly until this constraint is satisfied.

The integration can be simplified by selecting $a_\mathrm{ini}$ in the radiation era.
There $H \propto a^{-2}$ and $\bar\rho - 3 \bar{P} \simeq \bar\rho_r - 3 \bar{P}_r = 0$,
so the Klein-Gordon equation \eqref{eq:scalar_field_dimensionful} gives
$\smash{\dot{\bar\phi}} \propto a^{-3}$ with the solution $\bar\phi = A/a + B$.
The diverging solution is unphysical, leaving a frozen $\bar\phi = \mathrm{const}$ with
\begin{equation}
    \smash{\dot{\bar\phi}} = \smash{\dot{\bar\phi}_\mathrm{ini}} = 0 \qquad \text{(during radiation domination)}.
\label{eq:scalar_field_radiation_domination}
\end{equation}
In the Poisson equation \eqref{eq:poisson} below, we see that
the effective gravitational strength felt by matter today depends on $\omega$ and $\bar\phi_0$.
We therefore used the shooting method to vary $\bar\phi_\mathrm{ini}$ and integrate the system repeatedly until we hit the desired $\bar\phi_0$.

\subsection{Linear perturbations}
\label{sec:perturbations}

Ignoring anisotropic relativistic stresses and polarization
and working in the quasi-static limit $\{ \dot{\Phi}, \dot{\Psi}, \dot{\delta\phi} \} = H \cdot \{ \Phi, \Psi, \delta\phi \}$ and sub-horizon limit $k \gg H$
\citep{orjuela-quintanaTrackingValidityQuasistatic2023},
the linearly perturbed Boltzmann, Einstein and Klein-Gordon equations in $k$-space are
\citep{solaBransDickeGravityCosmological2019}
\begin{subequations}
\begin{align}
(\delta_c + 3\Phi)^\prime &= \phantom{-} \frac{k}{aH} v_c, \label{eq:pertc} \\
(\delta_b + 3\Phi)^\prime &= \phantom{-} \frac{k}{aH} v_b, \label{eq:pertb} \\
(\delta_\gamma + 4\Phi)^\prime &= \frac43 \frac{k}{aH} v_\gamma, \\
(a v_c)^\prime &= -\frac{k}{H} \Psi, \label{eq:pertvc} \\
(a v_b)^\prime &= -\frac{k}{H} \Psi + \frac43 \frac{\bar{\rho}_\gamma}{\bar{\rho}_b}\left(v_b - v_\gamma\right) a \tau^\prime , \label{eq:pertvb} \\
(a v_\gamma)^\prime &= -\frac{k}{H} \Psi - \frac{k}{4H} \delta_\gamma + av_\gamma + (v_\gamma-v_b) a \tau^\prime, \label{eq:pertvg} \\
k^2 \Big( \Phi + \frac12 \frac{\delta\phi}{\bar\phi} \Big) &= 4\pi \frac{G}{\bar\phi} a^2 \delta\rho, \label{eq:pertpoisson} \\
k^2 \frac{\delta\phi}{\bar\phi} &= \frac{8\pi}{2\omega+3} \frac{G}{\bar\phi} a^2 (\delta\rho - 3 \, \delta P), \label{eq:pertfieldevo} \\
\Psi + \Phi &= -\frac{\delta\phi}{\bar\phi}. \label{eq:pertfieldconstraint}
\end{align}%
\label{eq:perturbations}%
\end{subequations}
Motion of matter is sensitive to $\Psi$ through the geodesic equation.
In matter domination, $\delta P \simeq 0$ and $\delta \rho \simeq \bar\rho_m \delta_m \equiv \bar\rho_b \delta_b + \bar\rho_c \delta_c$,
and solving equations \eqref{eq:pertpoisson}, \eqref{eq:pertfieldevo} and \eqref{eq:pertfieldconstraint} for $\Psi$ gives its Poisson equation
\begin{equation}
    -k^2 \Psi = 4 \pi \underbrace{\frac{2\omega+4}{2\omega+3} \frac{G}{\bar\phi}}_{G_m} \frac{\bar\rho_{m0}}{a} \delta_m = \frac32 \frac{G_m}{G} \frac{\Omega_{m0} H_0^2}{a} \delta_m.
\label{eq:poisson}
\end{equation}
The Poisson equation in BD is the same as in GR,
only with an effective gravitational strength $G_m = (2\omega+4)/(2\omega+3) (G/\bar\phi) = (2\omega+4)/(2\omega+3) G_H$ felt by matter that depends on time through the scalar field in the background.
Cavendish experiments today measure $G_{m0}$, so the scalar field today must be $\bar\phi_0 = (2\omega+4)/(2\omega+3)$ in ``restricted models'' with the correct Newtonian limit $G_{m0} = G$.
However, we also allowed for ``unrestricted models'' with arbitrary $G_{m0}$ and $\bar\phi_0$, so we can constrain $G_{m0}$ with cosmology \citep{Zahn_2003,Ballardini_2022}.

To find a simplified growth equation for the total matter overdensity $\delta_m$,
baryon-photon interactions at late times can be neglected.
Combining the Boltzmann equations \eqref{eq:pertc}, \eqref{eq:pertb}, \eqref{eq:pertvc} and \eqref{eq:pertvb} with the Poisson equation \eqref{eq:poisson} then gives the growth equation
\begin{align}\label{eq:growth}
    0 &= \ddot{\delta}_m + 2H\dot{\delta}_m - 4 \pi G_m \bar\rho_m \delta_m \nonumber \\
      &= H^2 \bigg\{ \delta_m^{\prime\prime} + \Big[ 2 + (\ln E)^\prime \Big] \delta_m^\prime - \frac32 \frac{G_m}{G} \Omega_{m}(a) \delta_m \bigg\}.
\end{align}
The general solution of this second-order differential equation is a linear combination
$\delta_m(k,a) = C_+(k) D_+(a) + C_-(k) D_-(a)$
of a growing mode $D_+(a)$ and a decaying mode $D_-(a)$.
The growing mode is called the scale-independent growth factor $D_+(a)$,
and, neglecting the decaying mode, relates $D_+(a_2)/D_+(a_1) = \delta_{m}(k,a_2) / \delta_{m}(k,a_1)$ for any $k$.

On the other hand, deflection of light through weak lensing is sensitive to the Weyl potential $(\Psi-\Phi)/2$.
To find its source equation, eliminate $\delta\phi$ from equations \eqref{eq:pertpoisson}, \eqref{eq:pertfieldevo}, and \eqref{eq:pertfieldconstraint} to get
\begin{equation}
-k^2 \left(\frac{\Phi-\Psi}{2} \right) = 4 \pi \frac{G}{\bar\phi} \bar\rho_{m0} a^{-1} \delta_m.
\label{eq:lensing}
\end{equation}
Here we see that it is the effective $G_\gamma = G/\bar\phi$ that affects light trajectories.
In general, we encountered three effective gravitational strengths $G_\gamma = G_H \neq G_m$ that affect light trajectories, the expansion and matter clustering, respectively.
Although they generally take on two different values, we have $G_\gamma = G_H \approx G_m$ for (viable) models with appreciable $\omega$.

\begin{figure*}[ht!]
\begin{multicols}{2}
\centering
\includegraphics{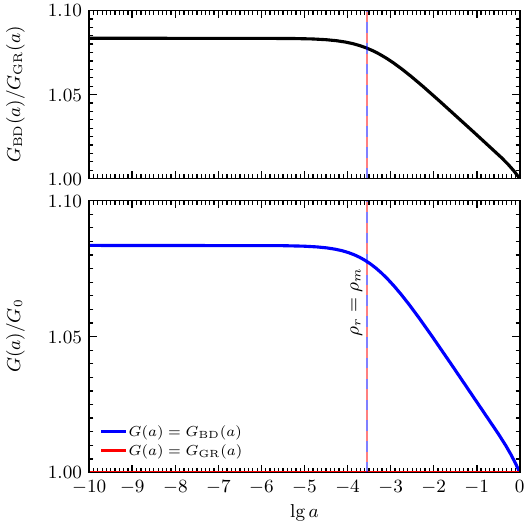}
\includegraphics{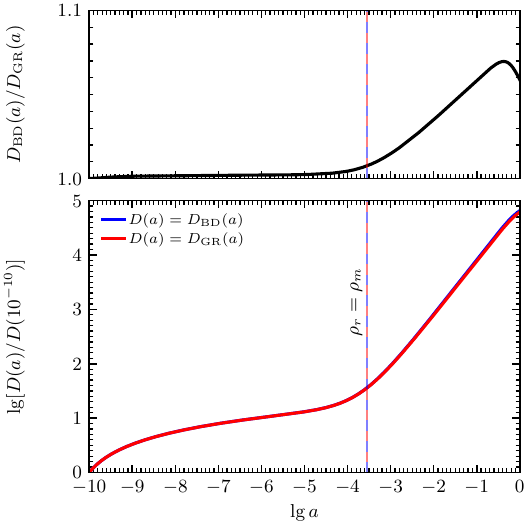}

\includegraphics{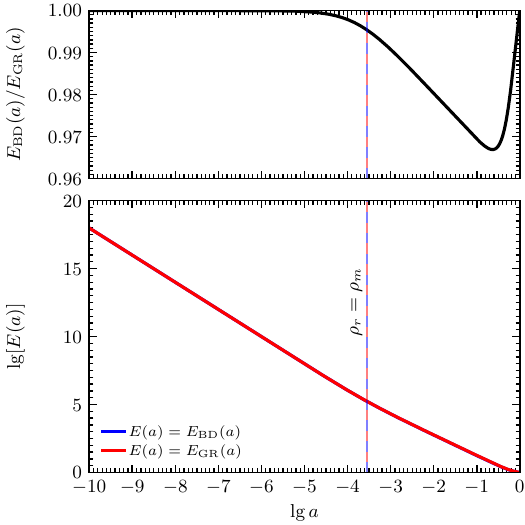}
\includegraphics{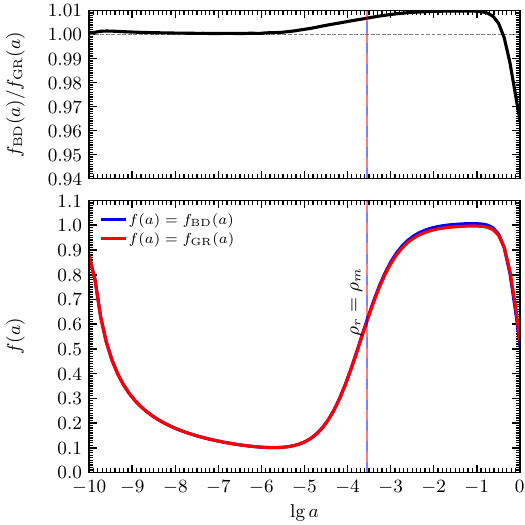}
\end{multicols}
\caption{Evolution of the effective gravitational parameter $G$, Hubble function $E(a) = H(a)/H_0$, growth factor $D(a)$, and growth rate $f(a) = \mathrm{d}\,{\ln D(a)}/\mathrm{d}\,{\ln a}$ from equations \eqref{eq:background_dimensionful} and \eqref{eq:growth} with the scale factor $a$ in the fiducial BD-$\LCDM$ and GR-$\LCDM$ cosmologies in \cref{tab:parameters} (equivalent to the transformation in \cref{fig:parametrization:s8_h2phi}). The common time of matter-radiation equality is marked by $\bar\rho_r = \bar\rho_m$.}
\label{fig:evolution_background}
\end{figure*}

The growth of structure is summarized by the matter power spectrum
\begin{subequations}
\begin{equation}
P(k,z) = P_\prim(k) \cdot |\Delta_m(k,z)|^2,
\label{eq:powerspectrum}
\end{equation}
of the gauge-invariant density perturbation $\Delta_m \approx \delta_m$
amplified from a nearly scale-invariant ($n_s \lesssim 1$) primordial power spectrum
\begin{equation}
    P_\prim(k) = \frac{2\pi^2}{k^3} \cdot A_s \left( \frac{k}{k_p} \right)^{n_s-1}
\label{eq:primordial}
\end{equation}
\end{subequations}
with primordial amplitude $A_s$ and spectral index $n_s$.

\subsection{Nonlinear structure formation}
To perform $N$-body simulations in BD we only need the geodesic equation
\begin{align}
    \frac{d^2\mathboldit{x}}{dt^2} + 2H\frac{d\mathboldit{x}}{dt} = -\frac{1}{a^2}\mathbold{\nabla} \Psi,
\label{eq:geodesic}
\end{align}
and the Poisson equation
\begin{align}
    \mathbold{\nabla}^2\Psi = \frac{3}{2}\Omega_{m0} \frac{G_m}{G} H_0^2 a^{-1}\delta_m.
\label{eq:poisson_nonlin}
\end{align}
The geodesic equation is the same in BD and GR, and the Poisson equation is the same as the linear version \eqref{eq:poisson}.
The only two modifications from a GR $N$-body simulation are the modified Hubble function $H(a)$ and the effective gravitational strength $G_m(a)$,
making it trivial to extend any $N$-body code from GR to BD.

\bigskip

\Cref{fig:evolution_background} compares the background evolution in two BD-$\LCDM$ and GR-$\LCDM$ universes with cosmological parameters to be introduced in \cref{sec:parametrization} and \cref{tab:parameters}.

\section{Computational pipeline}\label{sec:pipeline}

\begin{figure*}
\centering
\begin{tikzpicture}[
    node distance=0.5cm and 1.0cm,
    every node/.style = {
        draw, rectangle, rounded corners,
        text centered, align=center,
        minimum width=5.0cm, minimum height=1cm,
    },
]
\tikzstyle{arrow} = [thick,->,>=stealth]
\node [fill=black!10!white] {
    \begin{tikzpicture}
    \node (hiclass) [fill=blue!25!white] {\hiclass{}: \\ solve background; \\ solve perturbations. \\ };
    \node (fml)     [fill=blue!25!white, right=of hiclass] {\fml{}: \\ solve background; \\ seed particles with $s$; \\ evolve with COLA.};
    \node (ramses)  [fill=blue!25!white, right=of fml] {\ramses{}: \\ solve background; \\ read first snapshot; \\ evolve without COLA.};
    \node (par1)    [draw=none, above=of hiclass] {}; 
    \node (par2)    [draw=none, above=of fml]     {}; 
    \node (par3)    [draw=none, above=of ramses]  {}; 
    \node (phiclass)  [draw=none, below=of hiclass] {};
    \node (pfml)    [fill=red!10!white, below=of fml]     {Particle snapshots $\boldsymbol{x}_i(z_j)$};
    \node (pramses) [fill=red!10!white, below=of ramses]  {Particle snapshots $\boldsymbol{x}_i(z_j)$};
    \node (par)     [fill=green!26!white, above=of fml, xshift=0.0cm, minimum width=17.0cm] {Model parameters $\params$ and initial condition seed $s$};
    \node (Plin)    [fill=red!25!white, below=of phiclass] {Linear $P_\hiclass(k,z)$};
    \node (Pcola)       [fill=red!25!white, below=of pfml]  {Quasilinear $P_\mathrm{COLA}(k,z)$};
    \node (Pramses)     [fill=red!25!white, below=of pramses] {Nonlinear $P_\mathrm{RAMSES}(k,z)$};
    
    \draw [arrow] (par1)    to (hiclass);
    \draw [arrow] (par2)    to (fml);
    \draw [arrow] (par3)    to (ramses);
    \draw [arrow] (hiclass) to (Plin);
    \draw [arrow] (fml)     to (pfml);
    \draw [arrow] (pfml)    to (Pcola);
    \draw [arrow] (ramses)  to (pramses);
    \draw [arrow] (pramses)  to (Pramses);
    \draw [arrow] (Plin)    to [out=0, in=180, looseness=0.7] (fml);
    \draw [arrow] (pfml)    to [out=0, in=180, looseness=1.2] (ramses);

    \end{tikzpicture}
};
\end{tikzpicture}
\caption{\label{pipe:power_spectrum}%
    Our BD- or GR-specific simulation pipeline begins from a set of input parameters $\params_\BD$ or $\params_\GR$.
    First, they are passed to the Einstein-Boltzmann solver \hiclass{},
    which solves the background and perturbations
    and outputs the linear power spectrum.
    Next, \fml{} back-scales $P_\hiclass(k, z)$ from $z=0$ to $z=z_\mathrm{init}$,
    realizes initial conditions for an $N$-body simulation with the random seed $s$
    and evolves it forward with the COLA method on a uniform grid,
    outputting the nonlinear power spectrum.
    Finally, \fml{}'s first snapshot can be passed on to \ramses{},
    which evolves the same initial conditions with the standard $N$-body method on an adaptive mesh and yields an independent nonlinear power spectrum.
}
\end{figure*}
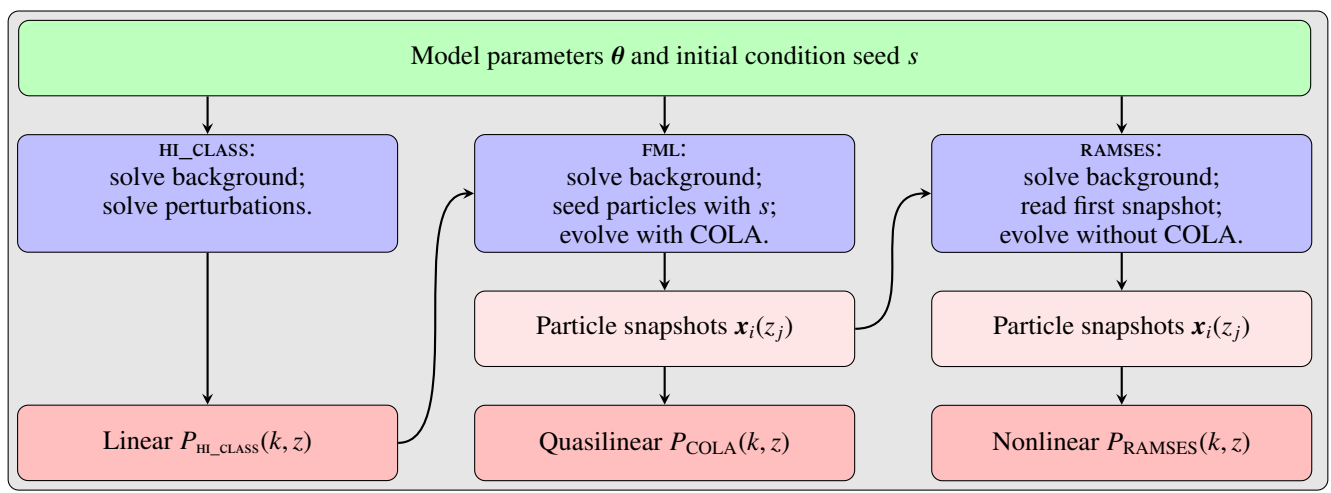

To compute the matter power spectrum $P(k,z\,|\,\params)$ in a BD-$\LCDM$ or GR-$\LCDM$ universe,
we joined three codes into the pipeline in \cref{pipe:power_spectrum}:
\begin{itemize}
\item
\hiclass{} \citep{zumalacarreguiHiClassHorndeski2017,belliniHiClassBackground2020},
based on \textsc{class} \citep{blasCosmicLinearAnisotropy2011},
solves the background and perturbation Einstein-Boltzmann equations
in GR-$\LCDM$ and several MG cosmologies, including BD-$\LCDM$.
The perturbed equations it solves are more sophisticated than our simplified presentation \eqref{eq:perturbations} and too lengthy to reproduce here.
It outputs the linear matter power spectrum $P_\hiclass(k,z)$.
\cite{belliniComparisonEinsteinBoltzmannSolvers2018} showed that its BD matter power spectrum agrees to $0.1\%$ with four other codes.
\item
\fml{}\footnote{The \fml{} library is available at \url{https://github.com/HAWinther/FML/}, and the COLA $N$-body code in the subdirectory \href{https://github.com/HAWinther/FML/tree/master/FML/COLASolver}{\nolinkurl{FML/COLASolver}}.} is a particle-mesh {\it N}-body code that uses the COLA method \citep{tassevSolvingLargeScale2013} in GR-$\LCDM$ and several MG cosmologies, including BD-$\LCDM$.
It supersedes the code \textsc{mg}-\textsc{picola} \citep{winther_cola_2017}, in turn based on \textsc{l}-\textsc{picola} \citep{howlett_l-picola:_2015}.
The code first backscales the linear power spectrum (see e.g. \cite{fidlerRelativisticInitialConditions2017}) from \hiclass{}, and then seeds initial particle positions.
It then solves the Poisson equation \eqref{eq:poisson_nonlin}
with the time-dependent $G_m(t)$ in BD and Newton's constant $G_m = G$ in GR,
Finally, it outputs the (quasi-)nonlinear matter power spectrum $P_\COLA(k,z)$ by constructing the density field from the particle position snapshots.
These results are used for our main analysis.
We used a box with volume $L^3 = (384\,h/\mathrm{Mpc})^3$ and $N_\mathrm{part} = N_\mathrm{cell} = 512^3$ particles and cells, evolving from redshift $z_\mathrm{init} = 10$ until today with $N_\mathrm{step} = 30$ time steps.
These simulations sacrifice accuracy at the deepest nonlinear scales for overall computational speed:
the particle-mesh nature of the code does not resolve small-scale forces,
and the COLA method maintains accuracy up to quasilinear scales with few time steps. 
This is an excellent trade-off in our scenario.
To some extent, the error in $P$ at smaller scales even cancels after we compute the boost $P_\BD/P_\GR$.
\item 
\ramses{} \citep{teyssierCosmologicalHydrodynamicsAdaptive2002}
solves the Poisson equation with the standard $N$-body method with adaptive mesh refinement and time-stepping
in GR-$\LCDM$ and (with our modifications) BD-$\LCDM$.
We modified its GR version\footnote{This \ramses{} patch is available at \url{https://github.com/HAWinther/JordanBransDicke}.} to solve the same Poisson equation as \fml{}
using arbitrary splined functions for $H(a)$ and $G_m(a)$ over time, passed on from \hiclass{}.
It starts from the same initial particle positions at the same redshift $z_\mathrm{init}$ as \fml{},
uses the same box size $L$ and number of particles $N_\mathrm{part}$,
adaptively refines a base grid with the same size $N_\mathrm{cell}$,
but uses adaptive time stepping that ignores \fml{}'s $N_\mathrm{step}$.
The snapshots are analyzed in the same way as \fml{},
except that we quadrupled the number of grid cells along each dimension due to the finer resolution.
This also outputs the nonlinear matter power spectrum $P_\ramses(k,z)$,
and is only used to validate \fml{}'s results.
\end{itemize}
All codes branch to solve the equations specific to BD-$\LCDM$ or GR-$\LCDM$ at all times.
To compensate for slight code differences, such as time stepping and splining effects,
we normalized $P$ from \fml{} and \ramses{} to match that from \hiclass{} at the most linear common scale $k = 0.02\,h/\mathrm{Mpc}$.

\subsection{Boost computation}

To compute the matter power spectrum boost
\begin{equation}
    B(k_\BD, z \,|\, \params_\BD) = \frac{P_\BD(k_\BD, z \,|\, \params_\BD)}{P_\GR(k_\GR,z \,|\, \params_\GR)},
\label{eq:boost}
\end{equation}
our pipeline first runs one BD simulation with freely chosen parameters $\params_\BD$,
followed by one GR simulation with parameters $\params_\GR = \params_\GR(\params_\BD)$ decided by a transformation of the BD parameters.
It then compares the power spectra at equal redshifts $z$,
but generally different wavenumbers $k_\BD$ and $k_\GR = k_\GR(k_\BD)$,
which we justify in \cref{sec:parametrization}.

To minimize cosmic variance within each $(\BD,\GR)$ simulation pair,
it uses a common initial condition seed $s_\BD = s_\GR = s$ (for fixed $\params_\BD$ and $\params_\GR$).
But the seed $s = s(\params_\BD)$ is changed for every new set of parameters to avoid bias toward one particular configuration of the universe.
To further reduce cosmic variance, we also used amplitude-fixed initial conditions \citep{2016MNRAS.462L...1A,Villaescusa_Navarro_2018}.

\subsection{Tests and convergence analysis}

\hiclass{} and \fml{} independently calculate $E(a)=H(a)/H_0$, $\Omega_{\Lambda 0}$, and $\sigma_8$ in GR, and also $\bar\phi(a)$ and $\bar\phi^\prime(a)$ in BD.
We verified that these agree within a small tolerance after every run.

\begin{figure*}
    \begin{multicols}{2}
    \centering
    \includegraphics{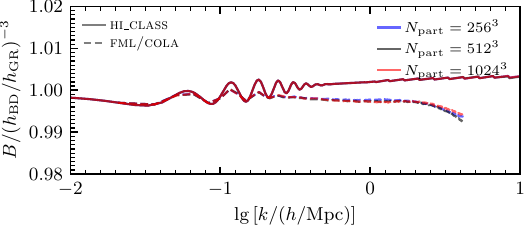}
    \includegraphics{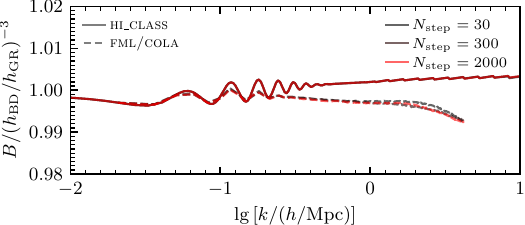}
    \includegraphics{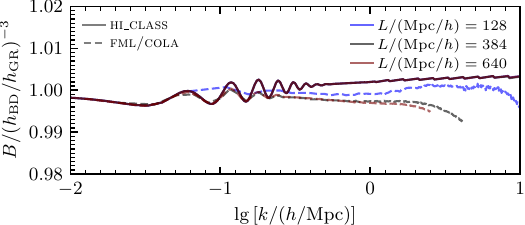}
    \includegraphics{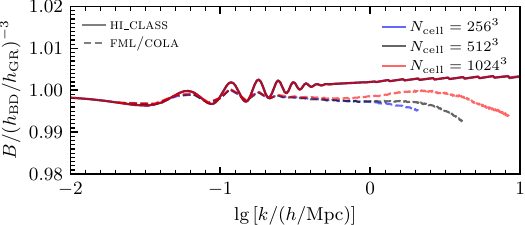}
    \includegraphics{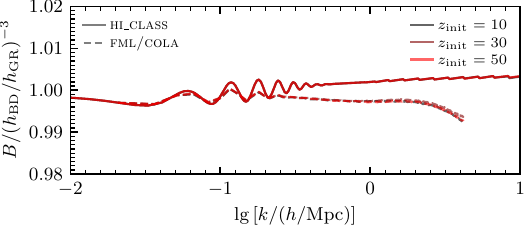}
    \end{multicols}
    \caption{Convergence of \fml{}'s boost \eqref{eq:boost} at $z=0$ in the fiducial cosmology in \cref{tab:parameters} (equivalent to the transformation in \cref{fig:parametrization:s8_h2phi}), as computational parameters are varied from the fiducial $N_\mathrm{part}=512^3$, $N_\mathrm{cell} = 512^3$, $N_\mathrm{step}=30$, $z_\mathrm{init}=10$, and $L=384\,\mathrm{Mpc}/h$.}
    \label{fig:convergence}
\end{figure*}

\begin{figure*}
    \begin{multicols}{2}
    \centering
    \includegraphics{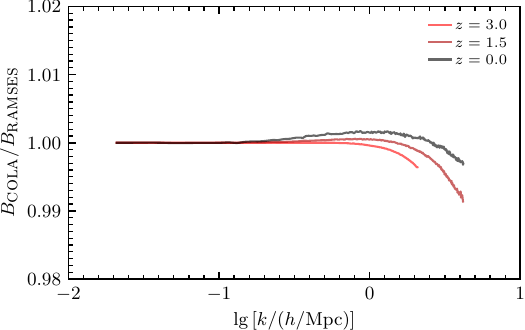} 
    
    \includegraphics{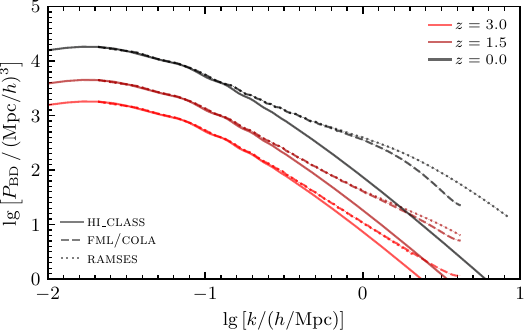}
    \end{multicols}
    \caption{Comparison of the matter power spectrum (boost) obtained from \fml{} using the COLA method on a uniform grid and \ramses{} using the standard $N$-body method on an adaptive grid, in the fiducial cosmology in \cref{tab:parameters} (equivalent to the transformation in \cref{fig:parametrization:s8_h2phi}). To compensate for small code differences, we have rescaled all spectra to match the linear \hiclass{} spectrum at their smallest common $k$. We cut \ramses{}' spectra at increasingly higher $k \leq \{1/2, 1, 2\} \cdot k_\mathrm{Nyquist}$ as the mesh refinement increases for $z = \{3,1.5,0\}$. Although $P$ from \fml{} and \ramses{} disagree, that error largely cancels in their boosts $B = P_\BD/P_\GR$, which agree to $1\%$ up to $k \lesssim 5\,h/\mathrm{Mpc}$.}
    \label{fig:ramses}
\end{figure*}

\Cref{fig:convergence} shows a convergence analysis of the results from \fml{}.
The results change by less than $1\%$ when the increasing the resolution of the simulation parameters,
showing that the results from our setup have converged.

To test the absolute accuracy of the particle-mesh simulations we ran using \fml{}, we also compared it to results from the adaptive-mesh-refinement code \ramses{}. These results are shown in \cref{fig:ramses} and again, the results are equal up to $1\%$ error up to $k \lesssim 5\,h/\mathrm{Mpc}$.
Together, these tests show that the results of the boost from \fml{} are both precise and accurate to $1\%$ error.
We only performed these tests for the fiducial cosmology in \cref{tab:parameters}:
we chose a small value $\omega=100$ in BD to maximize the deviation from GR, and assume the results can be trusted for other cosmological parameters.

\section{Model parameterization}
\label{sec:parametrization}

What parameter and wavenumber transformations $\params_\GR(\params_\BD)$ and $k_\GR(k_\BD)$ should we use in the boost \eqref{eq:boost}?
That is, how should we compare a BD universe to a GR universe?
The answer may sound as obvious as the question sounds pedantic: that we should use
the same parameters in BD and GR.
But this is vague or impossible, as it depends on our selection of cosmological parameters, and all cosmological parameters cannot be equal in two different universes.
For example, the background equations \eqref{eq:background_dimensionful}
forbid having the same $\bar\rho_{r0}$, $\bar\rho_{m0}$, $\bar\rho_{\Lambda 0}$, $H_0$ and $\bar\phi_0$ today.
For us this is a question of convenience.
We can choose to compare two universes in the boost \eqref{eq:boost}
such that this ratio behaves as simple and predictable as possible.
At the end of the day, we want to multiply it by $P_\GR$ to predict $P_\BD$, anyway.

\begin{figure*}
\begin{multicols}{2}
\centering
\subcaptionbox{Equal $\sigma_8=0.80$ and $\phi_\mathrm{ini} h^2 = 0.67$\label{fig:parametrization:s8_h2phi}}{\includegraphics{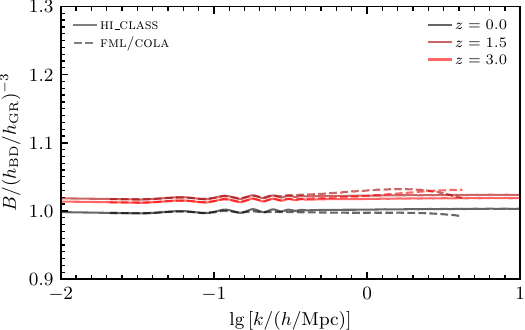}}
\\ \ \medskip \\
\subcaptionbox{Equal $A_s = 2 \cdot 10^{-9}$ and $\phi_\mathrm{ini} h^2 = 0.67$\label{fig:parametrization:As_h2phi}}{\includegraphics{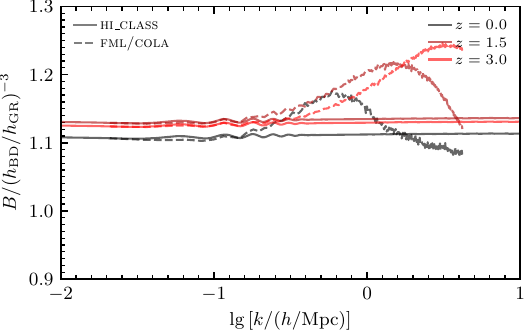}}

\subcaptionbox{Equal $\sigma_8=0.80$ and $h=0.7$\label{fig:parametrization:s8_h}}{\includegraphics{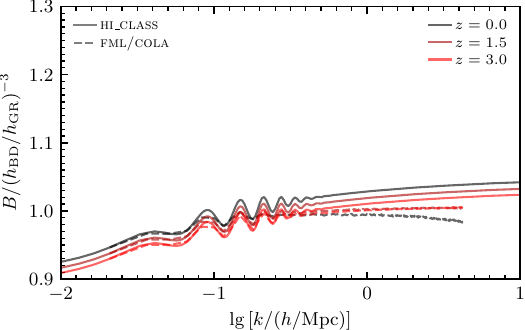}}
\\ \ \medskip \\
\subcaptionbox{Equal $A_s = 2 \cdot 10^{-9}$ and $h = 0.7$\label{fig:parametrization:As_h}}{\includegraphics{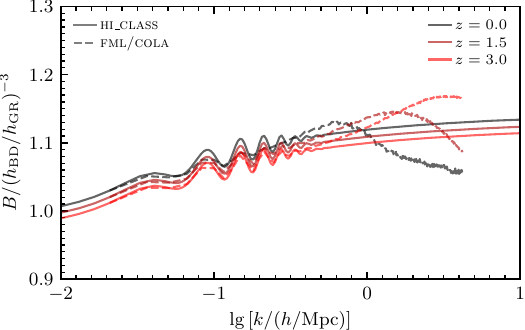}}
\end{multicols}
\caption{Matter power spectrum boost \eqref{eq:boost}
with different parameter transformations $\params_\GR = \params_\GR(\params_\BD)$,
where ($A_s^\BD = A_s^\GR$ or $\sigma_8^\BD = \sigma_8^\GR$) and ($H_0^\BD = H_0^\GR$ or $H_0^\GR = H_0^\BD \bar\phi_\mathrm{ini}^{1/2}$).
Requiring equal $\sigma_8$ (instead of equal $A_s$) flattens the boost in the nonlinear regime,
while requiring equal $\phi_\mathrm{ini} h^2$ (instead of equal $h$) synchronizes the sound horizon and avoids rapid baryon acoustic oscillations.
}
\label{fig:parametrization}
\end{figure*}

Next, we discuss some transformation alternatives while testing them in \cref{fig:parametrization}.

\subsection{Wavenumbers and Hubble parameter}
\label{sec:parametrization:wavenumbers_hubble}
Consider the perturbations \eqref{eq:perturbations} and forget about recombination for a moment,
which enters through the optical depth $\tau$.
In the Boltzmann equations,
$k$ and $H$ appear in the particular combination $k/H$.
At early times, the same is true for the Einstein and Klein-Gordon equations after substituting the effective $G_H = G/\bar\phi = 3 H^2 / 8 \pi \bar\rho$ from the Friedmann equation \eqref{eq:friedmann_dimensionful} with the frozen scalar field \eqref{eq:scalar_field_radiation_domination}.
For large enough $\omega$ that sufficiently suppress the anisotropic stress $\delta\phi/\bar\phi$,
perturbations in BD and GR with equal $k/H$ (or $k/\sqrt{G_H}$) evolve similarly.

In other words, BD resembles GR with a different expansion rate $H$
due to a different effective gravitational strength.
This difference can be factored out by comparing modes with equal $k/H$ (or $k/\sqrt{G_H}$).
Similarly, \cite[equation (12)]{Zahn_2003} shows that a mode $k'$ in a GR universe with gravitational constant $G'$ evolves like a mode $k = k'/\sqrt{G'/G}$ in another GR universe with Newton's constant $G$.
In other words, as they elegantly explained it,
``gravity introduces no preferred scale, so the dynamics of the perturbations remains the same when scales are measured in units of the expansion time.''
This also holds for BD's scale-independent effective gravitational strength.

However, it breaks when recombination is activated and $\tau$ is included,
as the faster expansion hinders (re)combination of $H^+ + e^-$
and dampens the small-scale acoustic peaks in the CMB spectrum.
This is indeed how \cite{Zahn_2003} shows how the CMB constrains $G$ to percent-level.
However, most matter is dark,
so this effect is much smaller for the power spectrum of matter that concerns us here.

This motivates matching modes $\{k_\BD, k_\GR\}$ such that $k/H$ is the same at early times.
To be compatible with $N$-body codes that factor $H_0$ out of the relevant equations,
we can compare modes with equal $k/H_0$ as long as we make $E_\BD = E_\GR$ at early times.
The frozen scalar field \eqref{eq:scalar_field_radiation_domination} cuts the Friedmann equation \eqref{eq:friedmann_dimensionful} off at $H_\BD^2 = H_\GR^2/\bar\phi$ provided we use the same physical densities $\bar\rho_{s0} \propto \omega_{s0}$ in BD and GR,%
\footnote{Strictly speaking, we cannot have $\bar\rho_{s0}^\GR =\bar\rho_{s0}^\BD$ for all species $s$ because $\bar\rho_{\Lambda 0}$ is constrained by the Friedmann equation, but we have $\bar\rho_\GR \approx \bar\rho_\BD$ at early times, when $\bar\rho_r \gg \bar\rho_m \gg \bar\rho_\Lambda$.}
so we can accomplish this by transforming
\begin{equation}
H_0^\GR = H^\BD_0 \bar\phi_\mathrm{ini}^{1/2}
\quad \iff \quad
H_0^\GR / \sqrt{\smash[b]{G_\GR^\mathrm{ini}}} = H_0^\BD / \sqrt{\smash[b]{G_\BD^\mathrm{ini}}}.
\label{eq:hubbletrans}
\end{equation}
This changes the simulation box size $L = 384\,\mathrm{Mpc}/h$ with $H_0 = 100\,h\,\mathrm{km}/\mathrm{s}/\mathrm{Mpc}$, so we can match modes at different scales.
Our ``scale scaling'' $k_\GR = (H_0^\GR/H_0^\BD) k_\BD = \bar\phi_\mathrm{ini}^{1/2} k_\BD$ contributes to our goal of ``mapping'' a BD universe onto a GR universe.

Another insightful way to understand this rescaling is to consider the sound horizon
\begin{equation}
    s = \int_0^t c_s \frac{\mathrm{d}t}{a} = \int_0^a \sqrt{\frac{4\bar\rho_\gamma/3\bar\rho_m}{1+4\bar\rho_\gamma/3\bar\rho_m}} \frac{\mathrm{d}a}{H},
\end{equation}
associated with baryonic acoustic oscillations (BAO).
This length scale ``freezes in'' when the baryons and photons decouple at recombination,
leaving a signature in the matter power spectrum at (multiples of) the characteristic wavenumber $k_s \propto 1/s$.
Indeed, the Hubble parameter transformation \eqref{eq:hubbletrans} ensures that $k_s/h$ will be the same in the two universes.
This synchronizes the phases of the oscillations in $P_\BD(k_\BD)$ and $P_\GR(k_\GR)$,
avoiding oscillations in their ratio $P_\BD/P_\GR$.
\Cref{fig:parametrization} shows the suppressed oscillations
when parameterized by $\bar\phi_\mathrm{ini} h^2$ instead of $h$.

The key to understand is that matching modes by $k/h$ combined with the parameter transformation \eqref{eq:hubbletrans} leverages the frozen scalar field \eqref{eq:scalar_field_radiation_domination} during radiation domination to make the initial evolution of perturbations as similar as possible in BD and GR.
This only holds until matter domination, when the scalar field begins to move.
But modes that have entered the horizon by this time simply grow (approximately) with the scale-independent growth factor.
These are precisely the interesting smaller-scale modes beyond the peak of the matter power spectrum!
Larger-scale modes are perfectly described by linear perturbation theory, anyway.

\subsection{Power spectrum normalization}
\label{sec:parametrization:power}
It is common to parameterize the primordial power spectrum \eqref{eq:primordial} with $A_s$.
However, as we are interested in predicting the late-time matter power spectrum,
it is more natural to ask for ``equal power'' in BD and GR today than at early times.
The normalization of today's power spectrum is conventionally done in terms of the parameter
\begin{equation}
    \sigma_R^2 = \frac{1}{2\pi^2} \int_0^\infty P_0(k) \hat{W}_R^2(k) k^2 \mathrm{d}k,
\label{eq:sigma}
\end{equation}
where $\hat{W}_R(k) = 3 \, [\sin(k R)-kR\cos(k R)] / (k R)^3$ is the Fourier transform of the top-hat profile
$W_R(r) = \Theta(R-r) / (4 \pi R^3 / 3)$
with radius $R$.
It is common to define the normalization using the value of $\sigma_8$, i.e. $\sigma_R$ with $R = 8\,\mathrm{Mpc}/h$.
By pinning $\sigma_8^\BD = \sigma_8^\GR$ today,%
\footnote{In principle, we must shoot to hit a given $\sigma_8(A_s)$.
But $A_s$ scales the whole linear power spectrum \eqref{eq:powerspectrum} and is detached from the linear perturbations, and the $\sigma_8$ integral \eqref{eq:sigma} dies off at nonlinear scales.
To avoid shooting, we can guess any value $A_s'$ of $A_s$, find $(\sigma_8')^2 \propto A_s'$ linearly, and then just proceed with $A_s = (\sigma_8 / \sigma_8')^2 A_s'$.}
we get less primordial power $A_s^\BD < A_s^\GR$
to compensate for the increased structure growth in BD
due to both less Hubble friction $E_\BD < E_\GR$ and stronger gravity $G_\BD > G_\GR$
(see \cref{fig:evolution_background}).
\Cref{fig:parametrization} shows that using the same $A_s$ gives a peak in the boost,
while using the same $\sigma_8$ gives a flatter boost.

\bigskip

This shows that the parameterization in \cref{fig:parametrization:s8_h2phi}
with $\sigma_8$ and $\bar\phi_\mathrm{ini} h^2$
gives the simplest and most predictable nonlinear boost.
As an added bonus, it also deviates the least from its linear counterpart,
and we show below that this holds through parameter space.

\begin{table}
    \centering
    \caption{Independent cosmological parameters $\params$ for BD-$\LCDM$ and GR-$\LCDM$ and, their fiducial values.}
    \begin{tabularx}{\columnwidth}{X r}
    \toprule
    Parameter & Fiducial value \\ 
    \midrule
    $\omega$ & $100$ \\ 
    $G_{m0}/G = \bar\phi_0^{-1} (2\omega+4)/(2\omega+3)$ & $1$ \\ 
    $h \, \bar\phi_\mathrm{ini}^{1/2}$ & $0.674$ \\ 
    $\sigma_8$ & $0.80$ \\ 
    $\omega_{m0} = \Omega_{m0} h^2$ & $0.15$ \\ 
    $\omega_{b0} = \Omega_{b0} h^2$ & $0.02$ \\ 
    $n_s$ & $1.0$ \\ 
    \bottomrule
    \end{tabularx}
    \tablefoot{In GR, $\omega = \infty$, $G_{m0}/G=1$ and $\bar\phi_\mathrm{ini} = 1$ implicitly. Listed parameters are varied, while the photon temperature today $T_{\gamma 0} = 2.7255\,\mathrm{K}$, effective massless neutrino number $N_\mathrm{eff} = 3.0$ and primordial power spectrum pivot scale $k_\mathrm{pivot} = 0.05\,\mathrm{Mpc}$ are fixed, and we always assume a flat universe. Derived parameters like the early scalar field $\bar\phi_\mathrm{ini}$, Hubble constant $h$, primordial power spectrum amplitude $A_s$ and dark energy density parameter $\omega_{\Lambda 0}$ can differ in BD and GR universes with equal independent parameters. Using the same values of these parameters in BD-$\LCDM$ and GR-$\LCDM$ is equivalent to using the parameter transformation in \cref{fig:parametrization:s8_h2phi}.}
    \label{tab:parameters}
\end{table}

\section{Resulting matter power spectrum boost}\label{sec:results}

We now focus on the boost with the particular parameterization developed in \cref{sec:parametrization},
defaulting to the fiducial parameters in \cref{tab:parameters} (equivalent to the transformation in \cref{fig:parametrization:s8_h2phi}).
In particular, due to the cosmological constraints $\omega \gtrsim 1000$ summarized in \cref{tab:constraints}, we restricted ourselves to $\omega \geq 100$ and used the minimum $\omega = 100$ in the fiducial cosmology.

\subsection{Understanding the linear boost growth}

The simplest way to understand the evolution of the linear scale-dependent boost is in terms of the scale-independent growth factor $D_+(a)$ from equation \eqref{eq:growth}.
During matter domination, the power spectrum scales like $P \propto \Delta_m^2 \approx \delta_m^2 \propto D_+^2$ at sub-horizon scales, so
\begin{equation}
    B(k,z) \propto \left( \frac{D_+^\BD(a(z))}{D_+^\GR(a(z))} \right)^2.
\label{eq:boost_scaleindependent}
\end{equation}

\begin{figure}[t]
    \centering
    \includegraphics{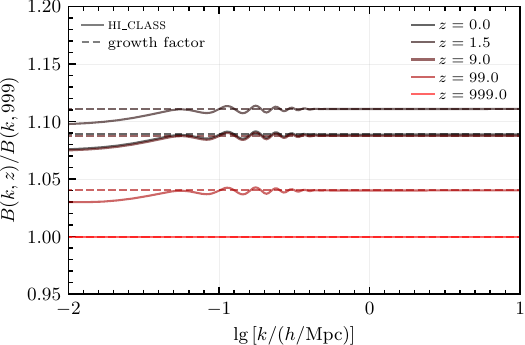}
    \caption{Evolution of the linear matter power spectrum boost from \hiclass{} compared to the scale-independent growth \eqref{eq:boost_scaleindependent} in \cref{fig:evolution_background}, starting from $a = 10^{-3}$ during early matter domination. The boosts are calculated in the fiducial cosmology in \cref{tab:parameters} (equivalent to the transformation in \cref{fig:parametrization:s8_h2phi}). Notice that the curves for $z=0$ and $z=9$ overlap,
    and that the boost turns around at the peak $z=1.5$ in \cref{fig:evolution_background}.}
    \label{fig:linear}
\end{figure}

\Cref{fig:linear} compares this to the evolution of the full linear boost from early matter domination till today.
As seen in \cref{fig:evolution_background}, the scalar field in BD begins to move during matter domination,
resulting in both less Hubble friction $E_\BD^\prime < E_\GR^\prime$ and stronger gravity $G_\BD > G_\GR$.
This gives faster growth in BD!

When dark energy takes over,
\cref{fig:evolution_background} shows that the expansion accelerates more in BD than in GR with $E_\BD^\prime > E_\GR^\prime$,
while the gravitational strengths $G_\BD \rightarrow G_\GR$ also become equal.
This causes the boost to peak around $z = 1.5$ and then fall off.

\subsection{Cosmological dependence of the nonlinear boost}

\begin{figure*}[h!]
\begin{multicols}{2}
\centering
\includegraphics{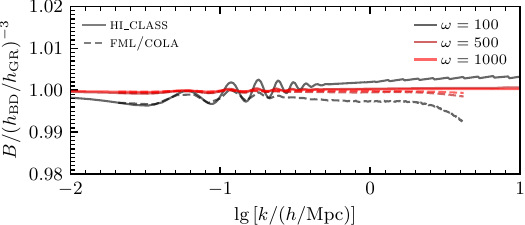}
\includegraphics{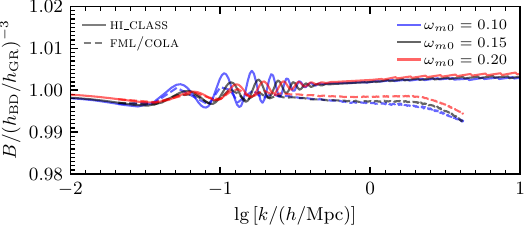}
\includegraphics{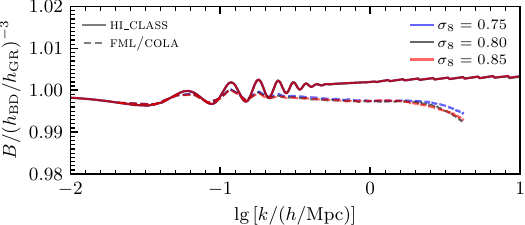}
\includegraphics{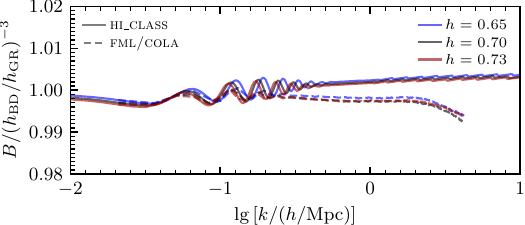}
\includegraphics{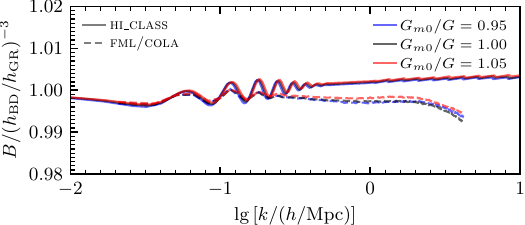}
\includegraphics{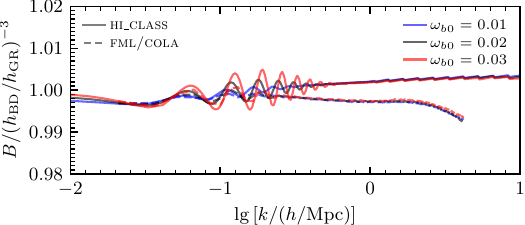}
\includegraphics{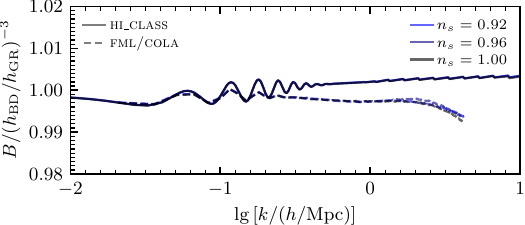}
\includegraphics{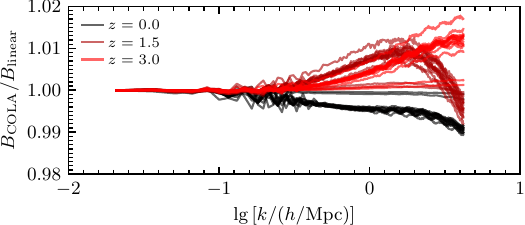}
\end{multicols}
\caption{Variation of the matter power spectrum boost \eqref{eq:boost} at $z=0$ as each cosmological parameter is varied away from its fiducial value in \cref{tab:parameters} (equivalent to the transformation in \cref{fig:parametrization:s8_h2phi}). The bottom right plot is special; showing the ratio of the nonlinear boost from \fml{}/COLA to the linear boost from \hiclass{} for all boosts in the 7 other plots at more redshifts $z=0.0,1.5,3.0$. It shows that $B \approx B_\lin$ to $1\%$ for $k \leq 1\,h/\mathrm{Mpc}$, and $2\%$ for $k \leq 5\,h/\mathrm{Mpc}$.}
\label{fig:variation}
\end{figure*}

We now study how the nonlinear boost changes with the cosmological parameters in \cref{tab:parameters}.
In \cref{fig:variation}, we vary one by one parameter away from its fiducial value while keeping the others fixed.
Not only is the boost (divided by $(h_\BD/h_\GR)^{-3} = \bar\phi_\mathrm{ini}^{3/2}$) very resistant toward parameter changes,
but so is its resemblance $P_\BD/P_\GR \approx P_\BD^\lin/P_\GR^\lin$ of the linear boost within $1\%$ up to $k \leq 1\,h/\mathrm{Mpc}$, and $2\%$ up to $k \leq 5\,h/\mathrm{Mpc}$.

\subsection{Linear prediction of the nonlinear boost}

Conveniently, this lets us get away with predicting the nonlinear boost $B \approx B_\lin=P_\BD^\lin/P_\GR^\lin$ using linear theory and codes, like \hiclass{}.
In turn, we can predict the nonlinear power spectrum $P_\BD = B_\lin \cdot P_\GR$
from an existing nonlinear prediction of $P_\GR$.
In fact, existing $P_\GR$ emulators like \eetwo{} only output the nonlinear correction factor
$C = P_\GR / P_\GR^\lin$ and defer multiplication with $P_\GR^\lin$,
so it even looks like we can cancel $P_\GR^\lin$
and simply calculate $P_\BD(\params_\BD) = C(\params_\GR(\params_\BD)) \cdot P_\BD^\lin(\params_\BD)$.
However, 
we still need to calculate $P_\GR^\lin$ in order to calculate the $\sigma_8$ integral \eqref{eq:sigma} from the perturbations to make the parameter transformation $\params_\GR(\params_\BD)$.

We provide the program \bdpy{}\footnote{The \bdpy{} code is available at \url{https://github.com/hersle/jbd}.} that handles these subtleties
and predicts the nonlinear BD spectrum 
from $2$ runs of \hiclass{} for the linear spectra BD and GR spectra
and $1$ run of \eetwo{} for the nonlinear GR spectrum.
This is our main product,
and can be used in future fitting to large-scale structure survey data.

\subsection{Comparison with existing nonlinear prediction}

\begin{figure*}
    \centering
    \begin{multicols}{2}
    \includegraphics{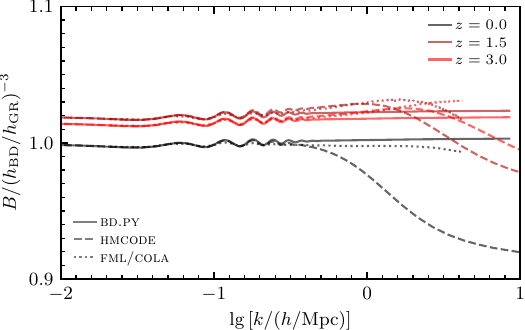}
    
    \includegraphics{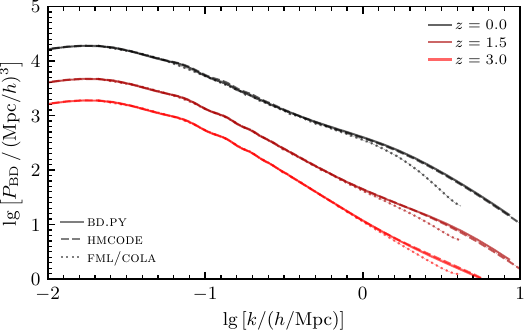}
    \end{multicols}
    \caption{A nonlinear BD-$\LCDM$ power spectrum (boost) with $\omega = 100$ obtained with our script \bdpy{} compared to that from an older fitting-formula modification to \hmcode{} in \hiclass{} from \cite{joudakiTestingGravityCosmic2022}, and COLA $N$-body results from \fml{}.}
    \label{fig:hmcode}
\end{figure*}

To demonstrate our program \bdpy{},
we compared it to the prediction of \cite{joudakiTestingGravityCosmic2022} that we mentioned in \cref{sec:intro}.
They incorporated BD into \hmcode{} by tuning the expression for the virial overdensity to match a set of $N$-body simulations. 
For their purpose of using \textit{KiDS} data to constrain the model, the accuracy aim was only a few percent up to $k\lesssim 1\, h/\mathrm{Mpc}$ and $5$-$10\%$ up to $k \lesssim 10\,h/\mathrm{Mpc}$.
\Cref{fig:hmcode} shows that their prediction was (almost) within their tolerance,
but produces too little power in BD when trialed with a tighter $1\%$ tolerance.

\section{Conclusion}\label{sec:conclusions}

Despite increasingly tight constraints from small-scale experiments,
BD gravity remains interesting at cosmological scales,
where constraints are weaker and it approximates more general theories of gravity with screening that falls back to GR at small scales.
More precise data from upcoming stage-IV large-scale structure surveys, like \textit{Euclid}, can increase the competitiveness of cosmological constraints,
and shed new light on this situation to illuminate the path forward for constraints on gravity.

In fact, this work began as an effort to train a traditional machine learning emulator for the nonlinear BD matter power spectrum from a set of COLA $N$-body simulations.
Instead, we found that it is possible to predict the nonlinear boost $B=P_\BD/P_\GR$ using linear theory and codes
if one transforms the cosmological parameters $\params_\BD \rightarrow \params_\GR$ in a particular way.
In detail, by comparing the linear boost with the nonlinear boosts from both COLA and standard $N$-body simulations, we verified that
\begin{gather*}
    \frac{P_\BD(k/h_\BD,z)}{P_\GR(k/h_\GR,z)}
    \overset{2\%}{\simeq}
    \frac{P_\BD^\mathrm{lin}(k/h_\BD,z)}{P_\GR^\mathrm{lin}(k/h_\GR,z)}
    \nonumber\\\text{for} \quad
    \begin{Bmatrix*}[l]
        k \lesssim 5\,h/\mathrm{Mpc} \\
        z \lesssim 3
    \end{Bmatrix*}
    \quad \text{if} \quad
    \begin{Bmatrix*}[l]
        \omega_\BD \gtrsim 100 \\
        \sigma_8^\GR = \sigma_8^\BD \\
        H_0^\GR = H_0^\BD \smash{\bar\phi_\mathrm{ini}^{1/2}} \\
        \bar\rho_{m0}^\GR = \bar\rho_{m0}^\BD \\
        \bar\rho_{r0}^\GR = \bar\rho_{r0}^\BD
    \end{Bmatrix*}.
\end{gather*}
This ``trick'' exploits the boost's dependence on cosmological parameters to map the nonlinear boost to the linear boost.

This has the advantage that the boost can be computed with cheap linear codes like \hiclass{},
instead of expensive nonlinear $N$-body methods.
When paired with an existing nonlinear $P_\GR$ predictor, such as \eetwo{},
this paves a simple and efficient path for predicting $P_\BD$ and constraining BD theory with the precision of upcoming stage-IV large-scale structure surveys. 
We provide the program \bdpy{} for this.

A drawback of our approach is that the parameter transformation $\params_\BD \rightarrow \params_\GR$ requires the emulator $P_\GR(\params_\GR)$ to cover a broader range of cosmological parameters than that motivated by GR alone.
For example, for reasonable input parameters $h_\BD$ and $A_s^\BD$, our transformation can request values of $h_\GR$ and $A_s^\GR$ that are outside the parameter ranges covered by a GR emulator.
To facilitate MG studies,
we therefore encourage developers of base GR emulators to ``think big'' when setting parameter bounds.
It would also be valuable to develop a traditional matter power spectrum emulator dedicated to BD,
or improve the precision of previous halo modeling approaches.

We also encourage others to investigate whether such simplifying parameter mappings $\params_\MG \rightarrow \params_\GR$ exist for other MG theories.
Our findings could be specific to BD gravity
or apply to more theories, perhaps with a similar scale-independent nature.
Even if one does not find $P_\MG(\params_\MG)/P_\GR(\params_\GR) \approx P_\MG^\lin(\params_\MG)/P_\GR^\lin(\params_\GR)$ ``exactly,''
investing thought in a clever parameter transformation may still pay off by simplifying the shape of $P_\MG/P_\GR$
so that one can obtain it from $P_\MG^\lin/P_\GR^\lin$ through a fitting formula or ease the training of emulators.
This reduces the effort needed to explore alternatives to $\LCDM$ with MG,
so one does not have to duplicate the effort gone into GR for every such alternative.

\begin{acknowledgements}
We thank the Research Council of Norway for
their support under grant no. 287772.
\end{acknowledgements}

\renewcommand{\eprint}[2][]{\href{https://arxiv.org/abs/#2}{\nolinkurl{arXiv:#2}}} 
\bibliographystyle{aa}
\bibliography{jbd}

\begin{thebibliography}{71}
\expandafter\ifx\csname natexlab\endcsname\relax\def\natexlab#1{#1}\fi

\bibitem[{{Abazajian} {et~al.}(2016){Abazajian}, {Adshead}, {Ahmed}, {Allen}, {Alonso}, {Arnold}, {Baccigalupi}, {Bartlett}, {Battaglia}, {Benson}, {Bischoff}, {Borrill}, {Buza}, {Calabrese}, {Caldwell}, {Carlstrom}, {Chang}, {Crawford}, {Cyr-Racine}, {De Bernardis}, {de Haan}, {di Serego Alighieri}, {Dunkley}, {Dvorkin}, {Errard}, {Fabbian}, {Feeney}, {Ferraro}, {Filippini}, {Flauger}, {Fuller}, {Gluscevic}, {Green}, {Grin}, {Grohs}, {Henning}, {Hill}, {Hlozek}, {Holder}, {Holzapfel}, {Hu}, {Huffenberger}, {Keskitalo}, {Knox}, {Kosowsky}, {Kovac}, {Kovetz}, {Kuo}, {Kusaka}, {Le Jeune}, {Lee}, {Lilley}, {Loverde}, {Madhavacheril}, {Mantz}, {Marsh}, {McMahon}, {Meerburg}, {Meyers}, {Miller}, {Munoz}, {Nguyen}, {Niemack}, {Peloso}, {Peloton}, {Pogosian}, {Pryke}, {Raveri}, {Reichardt}, {Rocha}, {Rotti}, {Schaan}, {Schmittfull}, {Scott}, {Sehgal}, {Shandera}, {Sherwin}, {Smith}, {Sorbo}, {Starkman}, {Story}, {van Engelen}, {Vieira}, {Watson}, {Whitehorn}, \& {Kimmy Wu}}]{abazajian_cmb-s4_2016}
{Abazajian}, K.~N., {Adshead}, P., {Ahmed}, Z., {et~al.} 2016, arXiv e-prints, arXiv:1610.02743

\bibitem[{{Acquaviva} {et~al.}(2005){Acquaviva}, {Baccigalupi}, {Leach}, {Liddle}, \& {Perrotta}}]{Acquaviva_2005}
{Acquaviva}, V., {Baccigalupi}, C., {Leach}, S.~M., {Liddle}, A.~R., \& {Perrotta}, F. 2005, \prd, 71, 104025

\bibitem[{{Alonso} {et~al.}(2017){Alonso}, {Bellini}, {Ferreira}, \& {Zumalac{\'a}rregui}}]{alonso_observational_2017}
{Alonso}, D., {Bellini}, E., {Ferreira}, P.~G., \& {Zumalac{\'a}rregui}, M. 2017, \prd, 95, 063502

\bibitem[{{Amirhashchi} \& {Yadav}(2020)}]{amirhashchiConstrainingExactBransDicke2020}
{Amirhashchi}, H. \& {Yadav}, A.~K. 2020, Physics of the Dark Universe, 30, 100711

\bibitem[{{Angulo} \& {Pontzen}(2016)}]{2016MNRAS.462L...1A}
{Angulo}, R.~E. \& {Pontzen}, A. 2016, \mnras, 462, L1

\bibitem[{{Angulo} \& {White}(2010)}]{angulo_one_2010}
{Angulo}, R.~E. \& {White}, S.~D.~M. 2010, \mnras, 405, 143

\bibitem[{{Angulo} {et~al.}(2021){Angulo}, {Zennaro}, {Contreras}, {Aric{\`o}}, {Pellejero-Iba{\~n}ez}, \& {St{\"u}cker}}]{2021MNRAS.507.5869A}
{Angulo}, R.~E., {Zennaro}, M., {Contreras}, S., {et~al.} 2021, \mnras, 507, 5869

\bibitem[{{Archibald} {et~al.}(2018){Archibald}, {Gusinskaia}, {Hessels}, {Deller}, {Kaplan}, {Lorimer}, {Lynch}, {Ransom}, \& {Stairs}}]{archibald_testing_2018}
{Archibald}, A.~M., {Gusinskaia}, N.~V., {Hessels}, J. W.~T., {et~al.} 2018, \nat, 559, 73

\bibitem[{{Aric{\`o}} {et~al.}(2021){Aric{\`o}}, {Angulo}, {Contreras}, {Ondaro-Mallea}, {Pellejero-Iba{\~n}ez}, \& {Zennaro}}]{2021MNRAS.506.4070A}
{Aric{\`o}}, G., {Angulo}, R.~E., {Contreras}, S., {et~al.} 2021, \mnras, 506, 4070

\bibitem[{{Aric{\`o}} {et~al.}(2022){Aric{\`o}}, {Angulo}, \& {Zennaro}}]{2021arXiv210414568A}
{Aric{\`o}}, G., {Angulo}, R.~E., \& {Zennaro}, M. 2022, Open Res Europe 2022, 1:152

\bibitem[{{Avilez} \& {Skordis}(2014)}]{avilezCosmologicalConstraintsBransDicke2014}
{Avilez}, A. \& {Skordis}, C. 2014, \prl, 113, 011101

\bibitem[{{Ballardini} {et~al.}(2020){Ballardini}, {Braglia}, {Finelli}, {Paoletti}, {Starobinsky}, \& {Umilt{\`a}}}]{Ballardini_2020}
{Ballardini}, M., {Braglia}, M., {Finelli}, F., {et~al.} 2020, \jcap, 2020, 044

\bibitem[{Ballardini {et~al.}(2022)Ballardini, Finelli, \& Sapone}]{Ballardini_2022}
Ballardini, M., Finelli, F., \& Sapone, D. 2022, \jcap, 2022, 004

\bibitem[{{Ballardini} {et~al.}(2016){Ballardini}, {Finelli}, {Umilt{\`a}}, \& {Paoletti}}]{Ballardini_2016}
{Ballardini}, M., {Finelli}, F., {Umilt{\`a}}, C., \& {Paoletti}, D. 2016, \jcap, 2016, 067

\bibitem[{Ballardini {et~al.}(2019)Ballardini, Sapone, Umiltà, Finelli, \& Paoletti}]{Ballardini_2019}
Ballardini, M., Sapone, D., Umiltà, C., Finelli, F., \& Paoletti, D. 2019, \jcap, 2019, 049–049

\bibitem[{{Bellini} {et~al.}(2018){Bellini}, {Barreira}, {Frusciante}, {Hu}, {Peirone}, {Raveri}, {Zumalac{\'a}rregui}, {Avilez-Lopez}, {Ballardini}, {Battye}, {Bolliet}, {Calabrese}, {Dirian}, {Ferreira}, {Finelli}, {Huang}, {Ivanov}, {Lesgourgues}, {Li}, {Lima}, {Pace}, {Paoletti}, {Sawicki}, {Silvestri}, {Skordis}, {Umilt{\`a}}, \& {Vernizzi}}]{belliniComparisonEinsteinBoltzmannSolvers2018}
{Bellini}, E., {Barreira}, A., {Frusciante}, N., {et~al.} 2018, \prd, 97, 023520

\bibitem[{{Bellini} {et~al.}(2020){Bellini}, {Sawicki}, \& {Zumalac{\'a}rregui}}]{belliniHiClassBackground2020}
{Bellini}, E., {Sawicki}, I., \& {Zumalac{\'a}rregui}, M. 2020, \jcap, 2020, 008

\bibitem[{{Bertotti} {et~al.}(2003){Bertotti}, {Iess}, \& {Tortora}}]{bertottiTestGeneralRelativity2003}
{Bertotti}, B., {Iess}, L., \& {Tortora}, P. 2003, \nat, 425, 374

\bibitem[{{Blas} {et~al.}(2011){Blas}, {Lesgourgues}, \& {Tram}}]{blasCosmicLinearAnisotropy2011}
{Blas}, D., {Lesgourgues}, J., \& {Tram}, T. 2011, \jcap, 2011, 034

\bibitem[{{Bose} {et~al.}(2020){Bose}, {Cataneo}, {Tr{\"o}ster}, {Xia}, {Heymans}, \& {Lombriser}}]{2020MNRAS.498.4650B}
{Bose}, B., {Cataneo}, M., {Tr{\"o}ster}, T., {et~al.} 2020, \mnras, 498, 4650

\bibitem[{{Brando} {et~al.}(2022){Brando}, {Fiorini}, {Koyama}, \& {Winther}}]{brandoEnablingMatterPower2022}
{Brando}, G., {Fiorini}, B., {Koyama}, K., \& {Winther}, H.~A. 2022, \jcap, 2022, 051

\bibitem[{{Brans} \& {Dicke}(1961)}]{bransMachPrincipleRelativistic1961a}
{Brans}, C. \& {Dicke}, R.~H. 1961, Physical Review, 124, 925

\bibitem[{{Casas} {et~al.}(2017){Casas}, {Kunz}, {Martinelli}, \& {Pettorino}}]{casas_linear_2017}
{Casas}, S., {Kunz}, M., {Martinelli}, M., \& {Pettorino}, V. 2017, Physics of the Dark Universe, 18, 73

\bibitem[{{Cataneo} {et~al.}(2019){Cataneo}, {Lombriser}, {Heymans}, {Mead}, {Barreira}, {Bose}, \& {Li}}]{Cataneo_2019}
{Cataneo}, M., {Lombriser}, L., {Heymans}, C., {et~al.} 2019, \mnras, 488, 2121

\bibitem[{{Clifton} {et~al.}(2012){Clifton}, {Ferreira}, {Padilla}, \& {Skordis}}]{clifton_modified_2012}
{Clifton}, T., {Ferreira}, P.~G., {Padilla}, A., \& {Skordis}, C. 2012, \physrep, 513, 1

\bibitem[{{DESI Collaboration}(2016)}]{desi_collaboration_desi_2016}
{DESI Collaboration}. 2016, arXiv e-prints, arXiv:1611.00036

\bibitem[{{Dicke}(1962)}]{dickeMachPrincipleInvariance1962a}
{Dicke}, R.~H. 1962, Physical Review, 125, 2163

\bibitem[{{Euclid Collaboration} {et~al.}(2021){Euclid Collaboration}, {Knabenhans}, {Stadel}, {Potter}, {Dakin}, {Hannestad}, {Tram}, {Marelli}, {Schneider}, {Teyssier}, {Fosalba}, {Andreon}, {Auricchio}, {Baccigalupi}, {Balaguera-Antol{\'\i}nez}, {Baldi}, {Bardelli}, {Battaglia}, {Bender}, {Biviano}, {Bodendorf}, {Bozzo}, {Branchini}, {Brescia}, {Burigana}, {Cabanac}, {Camera}, {Capobianco}, {Cappi}, {Carbone}, {Carretero}, {Carvalho}, {Casas}, {Casas}, {Castellano}, {Castignani}, {Cavuoti}, {Cledassou}, {Colodro-Conde}, {Congedo}, {Conselice}, {Conversi}, {Copin}, {Corcione}, {Coupon}, {Courtois}, {Da Silva}, {de la Torre}, {Di Ferdinando}, {Duncan}, {Dupac}, {Fabbian}, {Farrens}, {Ferreira}, {Finelli}, {Frailis}, {Franceschi}, {Galeotta}, {Garilli}, {Giocoli}, {Gozaliasl}, {Graci{\'a}-Carpio}, {Grupp}, {Guzzo}, {Holmes}, {Hormuth}, {Israel}, {Jahnke}, {Keihanen}, {Kermiche}, {Kirkpatrick}, {Kubik}, {Kunz}, {Kurki-Suonio}, {Ligori}, {Lilje}, {Lloro}, {Maino}, {Marggraf}, {Markovic}, {Martinet}, {Marulli},
  {Massey}, {Mauri}, {Maurogordato}, {Medinaceli}, {Meneghetti}, {Metcalf}, {Meylan}, {Moresco}, {Morin}, {Moscardini}, {Munari}, {Neissner}, {Niemi}, {Padilla}, {Paltani}, {Pasian}, {Patrizii}, {Pettorino}, {Pires}, {Polenta}, {Poncet}, {Raison}, {Renzi}, {Rhodes}, {Riccio}, {Romelli}, {Roncarelli}, {Saglia}, {S{\'a}nchez}, {Sapone}, {Schneider}, {Scottez}, {Secroun}, {Serrano}, {Sirignano}, {Sirri}, {Stanco}, {Sureau}, {Tallada Cresp{\'\i}}, {Taylor}, {Tenti}, {Tereno}, {Toledo-Moreo}, {Torradeflot}, {Valenziano}, {Valiviita}, {Vassallo}, {Viel}, {Wang}, {Welikala}, {Whittaker}, {Zacchei}, \& {Zucca}}]{euclidcollaborationEuclidPreparationIX2021}
{Euclid Collaboration}, {Knabenhans}, M., {Stadel}, J., {et~al.} 2021, \mnras, 505, 2840

\bibitem[{{Ezquiaga} \& {Zumalac{\'a}rregui}(2017)}]{ezquiaga_dark_2017}
{Ezquiaga}, J.~M. \& {Zumalac{\'a}rregui}, M. 2017, \prl, 119, 251304

\bibitem[{{Fidler} {et~al.}(2017){Fidler}, {Tram}, {Rampf}, {Crittenden}, {Koyama}, \& {Wands}}]{fidlerRelativisticInitialConditions2017}
{Fidler}, C., {Tram}, T., {Rampf}, C., {et~al.} 2017, \jcap, 2017, 043

\bibitem[{{Fiorini} {et~al.}(2023){Fiorini}, {Koyama}, \& {Baker}}]{fiorini_fast_2023}
{Fiorini}, B., {Koyama}, K., \& {Baker}, T. 2023, \jcap, 2023, 045

\bibitem[{{Freire} {et~al.}(2012){Freire}, {Wex}, {Esposito-Far{\`e}se}, {Verbiest}, {Bailes}, {Jacoby}, {Kramer}, {Stairs}, {Antoniadis}, \& {Janssen}}]{freireRelativisticPulsarwhiteDwarf2012}
{Freire}, P. C.~C., {Wex}, N., {Esposito-Far{\`e}se}, G., {et~al.} 2012, \mnras, 423, 3328

\bibitem[{{Frusciante} {et~al.}(2023){Frusciante}, {Pace}, {Cardone}, {Casas}, {Tutusaus}, {Ballardini}, {Bellini}, {Benevento}, {Bose}, {Valageas}, {Bartolo}, {Brax}, {Ferreira}, {Finelli}, {Koyama}, {Legrand}, {Lombriser}, {Paoletti}, {Pietroni}, {Rozas-Fern{\'a}ndez}, {Sakr}, {Silvestri}, {Vernizzi}, {Winther}, {Aghanim}, {Amendola}, {Auricchio}, {Azzollini}, {Baldi}, {Bonino}, {Branchini}, {Brescia}, {Brinchmann}, {Camera}, {Capobianco}, {Carbone}, {Carretero}, {Castellano}, {Cavuoti}, {Cimatti}, {Cledassou}, {Congedo}, {Conversi}, {Copin}, {Corcione}, {Courbin}, {Cropper}, {Da Silva}, {Degaudenzi}, {Dinis}, {Dubath}, {Dupac}, {Dusini}, {Farrens}, {Ferriol}, {Fosalba}, {Frailis}, {Franceschi}, {Galeotta}, {Gillis}, {Giocoli}, {Grazian}, {Grupp}, {Guzzo}, {Haugan}, {Holmes}, {Hormuth}, {Hornstrup}, {Jahnke}, {Kermiche}, {Kiessling}, {Kilbinger}, {Kitching}, {Kunz}, {Kurki-Suonio}, {Ligori}, {Lilje}, {Lloro}, {Maiorano}, {Mansutti}, {Marggraf}, {Markovic}, {Marulli}, {Massey}, {Medinaceli}, {Meneghetti},
  {Meylan}, {Moresco}, {Moscardini}, {Munari}, {Niemi}, {Nightingale}, {Padilla}, {Paltani}, {Pasian}, {Pedersen}, {Percival}, {Pettorino}, {Polenta}, {Poncet}, {Popa}, {Raison}, {Rebolo}, {Renzi}, {Rhodes}, {Riccio}, {Romelli}, {Saglia}, {Sapone}, {Sartoris}, {Secroun}, {Seidel}, {Sirignano}, {Sirri}, {Stanco}, {Surace}, {Tallada-Cresp{\'\i}}, {Taylor}, {Tereno}, {Toledo-Moreo}, {Torradeflot}, {Valentijn}, {Valenziano}, {Vassallo}, {Verdoes Kleijn}, {Wang}, {Zacchei}, {Zamorani}, {Zoubian}, \& {Scottez}}]{fruscianteEuclidConstrainingLinearly}
{Frusciante}, N., {Pace}, F., {Cardone}, V.~F., {et~al.} 2023, arXiv e-prints, arXiv:2306.12368

\bibitem[{Giblin {et~al.}(2019)Giblin, Cataneo, Moews, \& Heymans}]{Giblin_2019}
Giblin, B., Cataneo, M., Moews, B., \& Heymans, C. 2019, \mnras, 490, 4826–4840

\bibitem[{{Heitmann} {et~al.}(2016){Heitmann}, {Bingham}, {Lawrence}, {Bergner}, {Habib}, {Higdon}, {Pope}, {Biswas}, {Finkel}, {Frontiere}, \& {Bhattacharya}}]{heitmannMiraTitanUniversePrecision2016}
{Heitmann}, K., {Bingham}, D., {Lawrence}, E., {et~al.} 2016, \apj, 820, 108

\bibitem[{{Horndeski}(1974)}]{horndeski_second-order_1974}
{Horndeski}, G.~W. 1974, International Journal of Theoretical Physics, 10, 363

\bibitem[{{Howlett} {et~al.}(2015){Howlett}, {Manera}, \& {Percival}}]{howlett_l-picola:_2015}
{Howlett}, C., {Manera}, M., \& {Percival}, W.~J. 2015, Astronomy and Computing, 12, 109

\bibitem[{{Joudaki} {et~al.}(2022){Joudaki}, {Ferreira}, {Lima}, \& {Winther}}]{joudakiTestingGravityCosmic2022}
{Joudaki}, S., {Ferreira}, P.~G., {Lima}, N.~A., \& {Winther}, H.~A. 2022, \prd, 105, 043522

\bibitem[{{Laureijs} {et~al.}(2011){Laureijs}, {Amiaux}, {Arduini}, {Augu{\`e}res}, {Brinchmann}, {Cole}, {Cropper}, {Dabin}, {Duvet}, {Ealet}, {Garilli}, {Gondoin}, {Guzzo}, {Hoar}, {Hoekstra}, {Holmes}, {Kitching}, {Maciaszek}, {Mellier}, {Pasian}, {Percival}, {Rhodes}, {Saavedra Criado}, {Sauvage}, {Scaramella}, {Valenziano}, {Warren}, {Bender}, {Castander}, {Cimatti}, {Le F{\`e}vre}, {Kurki-Suonio}, {Levi}, {Lilje}, {Meylan}, {Nichol}, {Pedersen}, {Popa}, {Rebolo Lopez}, {Rix}, {Rottgering}, {Zeilinger}, {Grupp}, {Hudelot}, {Massey}, {Meneghetti}, {Miller}, {Paltani}, {Paulin-Henriksson}, {Pires}, {Saxton}, {Schrabback}, {Seidel}, {Walsh}, {Aghanim}, {Amendola}, {Bartlett}, {Baccigalupi}, {Beaulieu}, {Benabed}, {Cuby}, {Elbaz}, {Fosalba}, {Gavazzi}, {Helmi}, {Hook}, {Irwin}, {Kneib}, {Kunz}, {Mannucci}, {Moscardini}, {Tao}, {Teyssier}, {Weller}, {Zamorani}, {Zapatero Osorio}, {Boulade}, {Foumond}, {Di Giorgio}, {Guttridge}, {James}, {Kemp}, {Martignac}, {Spencer}, {Walton}, {Bl{\"u}mchen}, {Bonoli},
  {Bortoletto}, {Cerna}, {Corcione}, {Fabron}, {Jahnke}, {Ligori}, {Madrid}, {Martin}, {Morgante}, {Pamplona}, {Prieto}, {Riva}, {Toledo}, {Trifoglio}, {Zerbi}, {Abdalla}, {Douspis}, {Grenet}, {Borgani}, {Bouwens}, {Courbin}, {Delouis}, {Dubath}, {Fontana}, {Frailis}, {Grazian}, {Koppenh{\"o}fer}, {Mansutti}, {Melchior}, {Mignoli}, {Mohr}, {Neissner}, {Noddle}, {Poncet}, {Scodeggio}, {Serrano}, {Shane}, {Starck}, {Surace}, {Taylor}, {Verdoes-Kleijn}, {Vuerli}, {Williams}, {Zacchei}, {Altieri}, {Escudero Sanz}, {Kohley}, {Oosterbroek}, {Astier}, {Bacon}, {Bardelli}, {Baugh}, {Bellagamba}, {Benoist}, {Bianchi}, {Biviano}, {Branchini}, {Carbone}, {Cardone}, {Clements}, {Colombi}, {Conselice}, {Cresci}, {Deacon}, {Dunlop}, {Fedeli}, {Fontanot}, {Franzetti}, {Giocoli}, {Garcia-Bellido}, {Gow}, {Heavens}, {Hewett}, {Heymans}, {Holland}, {Huang}, {Ilbert}, {Joachimi}, {Jennins}, {Kerins}, {Kiessling}, {Kirk}, {Kotak}, {Krause}, {Lahav}, {van Leeuwen}, {Lesgourgues}, {Lombardi}, {Magliocchetti}, {Maguire},
  {Majerotto}, {Maoli}, {Marulli}, {Maurogordato}, {McCracken}, {McLure}, {Melchiorri}, {Merson}, {Moresco}, {Nonino}, {Norberg}, {Peacock}, {Pello}, {Penny}, {Pettorino}, {Di Porto}, {Pozzetti}, {Quercellini}, {Radovich}, {Rassat}, {Roche}, {Ronayette}, {Rossetti}, {Sartoris}, {Schneider}, {Semboloni}, {Serjeant}, {Simpson}, {Skordis}, {Smadja}, {Smartt}, {Spano}, {Spiro}, {Sullivan}, {Tilquin}, {Trotta}, {Verde}, {Wang}, {Williger}, {Zhao}, {Zoubian}, \& {Zucca}}]{laureijsEuclidDefinitionStudy2011}
{Laureijs}, R., {Amiaux}, J., {Arduini}, S., {et~al.} 2011, arXiv e-prints, arXiv:1110.3193

\bibitem[{{Lawrence} {et~al.}(2017){Lawrence}, {Heitmann}, {Kwan}, {Upadhye}, {Bingham}, {Habib}, {Higdon}, {Pope}, {Finkel}, \& {Frontiere}}]{lawrenceMiraTitanUniverseII2017}
{Lawrence}, E., {Heitmann}, K., {Kwan}, J., {et~al.} 2017, \apj, 847, 50

\bibitem[{{Li} {et~al.}(2013){Li}, {Wu}, \& {Chen}}]{liConstraintsBransDickeGravity2013}
{Li}, Y.-C., {Wu}, F.-Q., \& {Chen}, X. 2013, \prd, 88, 084053

\bibitem[{{LSST Science Collaboration}(2009)}]{lsst_science_collaboration_lsst_2009}
{LSST Science Collaboration}. 2009, arXiv e-prints, arXiv:0912.0201

\bibitem[{{Mauland} {et~al.}(2024){Mauland}, {Winther}, \& {Ruan}}]{maulandSesamePowerSpectrum2023}
{Mauland}, R., {Winther}, H.~A., \& {Ruan}, C.-Z. 2024, \aap, 685, A156

\bibitem[{{Mead} {et~al.}(2021){Mead}, {Brieden}, {Tr{\"o}ster}, \& {Heymans}}]{mead_hmcode-2020:_2021}
{Mead}, A.~J., {Brieden}, S., {Tr{\"o}ster}, T., \& {Heymans}, C. 2021, \mnras, 502, 1401

\bibitem[{{Mead} {et~al.}(2016){Mead}, {Heymans}, {Lombriser}, {Peacock}, {Steele}, \& {Winther}}]{mead_accurate_2016}
{Mead}, A.~J., {Heymans}, C., {Lombriser}, L., {et~al.} 2016, \mnras, 459, 1468

\bibitem[{{Mead} {et~al.}(2015){Mead}, {Peacock}, {Heymans}, {Joudaki}, \& {Heavens}}]{mead_accurate_2015}
{Mead}, A.~J., {Peacock}, J.~A., {Heymans}, C., {Joudaki}, S., \& {Heavens}, A.~F. 2015, \mnras, 454, 1958

\bibitem[{{Moran} {et~al.}(2023){Moran}, {Heitmann}, {Lawrence}, {Habib}, {Bingham}, {Upadhye}, {Kwan}, {Higdon}, \& {Payne}}]{moranMiraTitanUniverseIV2023}
{Moran}, K.~R., {Heitmann}, K., {Lawrence}, E., {et~al.} 2023, \mnras, 520, 3443

\bibitem[{{Nagata} {et~al.}(2004){Nagata}, {Chiba}, \& {Sugiyama}}]{Nagata_2004}
{Nagata}, R., {Chiba}, T., \& {Sugiyama}, N. 2004, \prd, 69, 083512

\bibitem[{{Ooba} {et~al.}(2016){Ooba}, {Ichiki}, {Chiba}, \& {Sugiyama}}]{Ooba_2016}
{Ooba}, J., {Ichiki}, K., {Chiba}, T., \& {Sugiyama}, N. 2016, \prd, 93, 122002

\bibitem[{{Ooba} {et~al.}(2017){Ooba}, {Ichiki}, {Chiba}, \& {Sugiyama}}]{Ooba_2017}
{Ooba}, J., {Ichiki}, K., {Chiba}, T., \& {Sugiyama}, N. 2017, Progress of Theoretical and Experimental Physics, 2017, 043E03

\bibitem[{{Orjuela-Quintana} \& {Nesseris}(2023)}]{orjuela-quintanaTrackingValidityQuasistatic2023}
{Orjuela-Quintana}, J.~B. \& {Nesseris}, S. 2023, \jcap, 2023, 019

\bibitem[{{Perivolaropoulos} \& {Skara}(2022)}]{perivolaropoulosChallengesCDMUpdate2022a}
{Perivolaropoulos}, L. \& {Skara}, F. 2022, \nar, 95, 101659

\bibitem[{{Ruan} {et~al.}(2024){Ruan}, {Cuesta-Lazaro}, {Eggemeier}, {Li}, {Baugh}, {Arnold}, {Bose}, {Hern{\'a}ndez-Aguayo}, {Zarrouk}, \& {Davies}}]{ruan2023emulatorbased}
{Ruan}, C.-Z., {Cuesta-Lazaro}, C., {Eggemeier}, A., {et~al.} 2024, \mnras, 527, 2490

\bibitem[{{S{\'a}ez-Casares} {et~al.}(2024){S{\'a}ez-Casares}, {Rasera}, \& {Li}}]{saez-casares_e-mantis_2023}
{S{\'a}ez-Casares}, I., {Rasera}, Y., \& {Li}, B. 2024, \mnras, 527, 7242

\bibitem[{{Smith} {et~al.}(2003){Smith}, {Peacock}, {Jenkins}, {White}, {Frenk}, {Pearce}, {Thomas}, {Efstathiou}, \& {Couchman}}]{smith_stable_2003}
{Smith}, R.~E., {Peacock}, J.~A., {Jenkins}, A., {et~al.} 2003, \mnras, 341, 1311

\bibitem[{{Sol{\`a} Peracaula} {et~al.}(2019){Sol{\`a} Peracaula}, {G{\'o}mez-Valent}, {de Cruz P{\'e}rez}, \& {Moreno-Pulido}}]{solaBransDickeGravityCosmological2019}
{Sol{\`a} Peracaula}, J., {G{\'o}mez-Valent}, A., {de Cruz P{\'e}rez}, J., \& {Moreno-Pulido}, C. 2019, \apjl, 886, L6

\bibitem[{{Sol{\`a} Peracaula} {et~al.}(2020){Sol{\`a} Peracaula}, {G{\'o}mez-Valent}, {de Cruz P{\'e}rez}, \& {Moreno-Pulido}}]{solaBransDickeCosmologyLambda2020}
{Sol{\`a} Peracaula}, J., {G{\'o}mez-Valent}, A., {de Cruz P{\'e}rez}, J., \& {Moreno-Pulido}, C. 2020, Class. Quantum Grav., 37, 245003

\bibitem[{{Takahashi} {et~al.}(2012){Takahashi}, {Sato}, {Nishimichi}, {Taruya}, \& {Oguri}}]{takahashi_revising_2012}
{Takahashi}, R., {Sato}, M., {Nishimichi}, T., {Taruya}, A., \& {Oguri}, M. 2012, \apj, 761, 152

\bibitem[{{Tassev} {et~al.}(2013){Tassev}, {Zaldarriaga}, \& {Eisenstein}}]{tassevSolvingLargeScale2013}
{Tassev}, S., {Zaldarriaga}, M., \& {Eisenstein}, D.~J. 2013, \jcap, 2013, 036

\bibitem[{{Teyssier}(2002)}]{teyssierCosmologicalHydrodynamicsAdaptive2002}
{Teyssier}, R. 2002, \aap, 385, 337

\bibitem[{{Umilt{\`a}} {et~al.}(2015){Umilt{\`a}}, {Ballardini}, {Finelli}, \& {Paoletti}}]{Umilt__2015}
{Umilt{\`a}}, C., {Ballardini}, M., {Finelli}, F., \& {Paoletti}, D. 2015, \jcap, 2015, 017

\bibitem[{Villaescusa-Navarro {et~al.}(2018)Villaescusa-Navarro, Naess, Genel, Pontzen, Wandelt, Anderson, Font-Ribera, Battaglia, \& Spergel}]{Villaescusa_Navarro_2018}
Villaescusa-Navarro, F., Naess, S., Genel, S., {et~al.} 2018, The Astrophysical Journal, 867, 137

\bibitem[{{Voisin} {et~al.}(2020){Voisin}, {Cognard}, {Freire}, {Wex}, {Guillemot}, {Desvignes}, {Kramer}, \& {Theureau}}]{voisinImprovedTestStrong2020}
{Voisin}, G., {Cognard}, I., {Freire}, P.~C.~C., {et~al.} 2020, \aap, 638, A24

\bibitem[{{Will}(2014)}]{willConfrontationGeneralRelativity2014}
{Will}, C.~M. 2014, Living Reviews in Relativity, 17, 4

\bibitem[{{Winther} {et~al.}(2019){Winther}, {Casas}, {Baldi}, {Koyama}, {Li}, {Lombriser}, \& {Zhao}}]{wintherEmulatorsNonlinearMatter2019}
{Winther}, H.~A., {Casas}, S., {Baldi}, M., {et~al.} 2019, \prd, 100, 123540

\bibitem[{{Winther} {et~al.}(2017){Winther}, {Koyama}, {Manera}, {Wright}, \& {Zhao}}]{winther_cola_2017}
{Winther}, H.~A., {Koyama}, K., {Manera}, M., {Wright}, B.~S., \& {Zhao}, G.-B. 2017, \jcap, 2017, 006

\bibitem[{{Wu} \& {Chen}(2010)}]{Wu_2010}
{Wu}, F.-Q. \& {Chen}, X. 2010, \prd, 82, 083003

\bibitem[{{Wu} {et~al.}(2010){Wu}, {Qiang}, {Wang}, \& {Chen}}]{wu_cosmic_2010}
{Wu}, F.-Q., {Qiang}, L.-E., {Wang}, X., \& {Chen}, X. 2010, \prd, 82, 083002

\bibitem[{{Yahya} {et~al.}(2015){Yahya}, {Bull}, {Santos}, {Silva}, {Maartens}, {Okouma}, \& {Bassett}}]{yahya_cosmological_2015}
{Yahya}, S., {Bull}, P., {Santos}, M.~G., {et~al.} 2015, \mnras, 450, 2251

\bibitem[{{Zahn} \& {Zaldarriaga}(2003)}]{Zahn_2003}
{Zahn}, O. \& {Zaldarriaga}, M. 2003, \prd, 67, 063002

\bibitem[{{Zumalac{\'a}rregui} {et~al.}(2017){Zumalac{\'a}rregui}, {Bellini}, {Sawicki}, {Lesgourgues}, \& {Ferreira}}]{zumalacarreguiHiClassHorndeski2017}
{Zumalac{\'a}rregui}, M., {Bellini}, E., {Sawicki}, I., {Lesgourgues}, J., \& {Ferreira}, P.~G. 2017, \jcap, 2017, 019

\end{thebibliography}

\end{document}